\documentclass[a4paper,amsfonts, notitlepage,11pt,superscriptaddress,nofootinbib,floatfix,showpacs]{revtex4-1}
\pdfoutput=1
\usepackage[T1]{fontenc}
\usepackage[utf8]{inputenc}
\usepackage[english]{babel}
\usepackage{amsmath,amssymb,amsfonts}
\usepackage{mathrsfs}
\usepackage{bbm}
\usepackage[dvipdf,dvips,dvipdfmx]{graphicx}
\usepackage{verbatim}
\usepackage{setspace}
\usepackage[usenames,dvipsnames]{xcolor}
\usepackage{mathbbol}
\usepackage[colorlinks=true,linkcolor=RoyalBlue,citecolor=RoyalBlue,linktocpage=true]{hyperref}
\usepackage{multirow}
\usepackage[export]{adjustbox}

\makeatother

\begin{document}
\title{From renormalization group flows to cosmology}
\author{Alessia Platania}
\email{a.platania@thphys.uni-heidelberg.de}
\affiliation{Institut für Theoretische Physik, Universität Heidelberg, Philosophenweg 16, 69120 Heidelberg, Germany}

\begin{abstract}
According to the asymptotic-safety conjecture, the gravitational renormalization
group flow features an ultraviolet-attractive fixed point that makes
the theory renormalizable and ultraviolet complete. The existence
of this fixed point entails an antiscreening of the gravitational
interaction at short distances. In this paper we review the state-of-the-art
of phenomenology of Asymptotically Safe Gravity, focusing on the implications
of the gravitational antiscreening in cosmology.
\end{abstract}
\maketitle

\tableofcontents{}

\section{Gravitational antiscreening: a historical perspective}

Similarly to the case of Quantum Chromodynamics (QCD), the gravitational
interaction might exhibit an antiscreening behavior at high energies~\cite{Nink:2012vd}. A form of gravitational antiscreening was introduced in the 80s by Markov~\cite{1982JETPL..36..265M,1983PhLA...94..427M} as a mechanism to cure the longstanding problem of gravitational singularities
in General Relativity. In~\cite{Markov:1985py,Markov:1994hi} an \emph{ad
hoc} modification of the Einstein-Hilbert Lagrangian has been proposed,
such that the corresponding field equations
\begin{equation}
G_{\mu\nu}=G(\rho)\,T_{\mu\nu}+\Lambda(\rho)\,g_{\mu\nu}\, ,
\end{equation}
admit a Newton coupling $G(\rho)$ and cosmological constant $\Lambda(\rho)$
whose strengths depend on the proper energy density $\rho$ of matter
fields, with $G(\rho)\to0$ as $\rho\to\infty$. This latter assumption,
which in~\cite{Markov:1985py,Markov:1994hi} is referred to as ``asymptotic
freedom of gravity'', has been introduced to render the gravitational
interaction weaker at high energy densities. In a cosmological context,
this could lead to a singularity-free cosmological evolution characterized
by a deSitter initial state and a minimum radius of the order of the
Planck length, $a_{min}\sim L_{Pl}$ . In this case the resolution
of the cosmological singularity is due to the violation of the energy
conditions, thus invalidating one of the key assumptions leading to
the Hawking-Penrose singularity theorems~\cite{1970HP}. Similar arguments
could also apply to the case of black holes, where the gravitational
antiscreening could lead to singularity-free black-hole spacetimes
\cite{Frolov:1989pf}.

The ``asymptotic freedom of gravity'' discussed in~\cite{Markov:1985py,Markov:1994hi}
was originally introduced as a modification of General Relativity
at the classical level. It turns out that the gravitational antiscreening
advocated in~\cite{Markov:1985py,Markov:1994hi} could be a natural
consequence of the quantum properties of gravity. The ``asymptotic
safety'' scenario for Quantum Gravity~\cite{Reuter:2012id,Percacci:2017fkn,Reuter:2019byg}
aims at constructing a consistent quantum theory for the gravitational
interaction within the well-established framework of Quantum Field
Theory (QFT). As originally proposed by Weinberg~\cite{1976W,1979W},
in the light of the Wilsonian renormalization group~\cite{Wilson1,Wilson2}
and the related, generalized notion of renormalizability~\cite{Wilson:1973jj},
a consistent QFT of gravity could be constructed if the gravitational
renormalization group (RG) flow attains an interacting -- non-Gaussian
-- fixed point (NGFP) in the ultraviolet limit. In this case, in
the ultraviolet regime, gravity approaches a scale invariant regime
where the dimensionless counterparts of all gravitational couplings
attain finite, generally non-zero, values. The theory is thus interacting
in the ultraviolet regime and the presence of the NGFP ensures the
``non-perturbative'' renormability~\cite{Wilson:1973jj} of gravity.
While asymptotic safety of gravity remains a formally unproved
conjecture, there are strong indications that a suitable gravitational
fixed point indeed exists~\cite{Reuter:1996cp,Souma:1999at,Reuter:2001ag,Litim:2003vp,Codello:2007bd,Benedetti:2009gn,Groh:2011vn,Donkin:2012ud,Benedetti:2013jk,Eichhorn:2013xr,Falls:2014tra,Demmel:2015oqa,Eichhorn:2015bna,Biemans:2016rvp,Gies:2016con,Hamada:2017rvn,Biemans:2017zca,Platania:2017djo,Falls:2018ylp}. Moreover, despite its interacting
nature, the fixed point could lie in the vicinity of the free-theory
fixed point (Gaussian fixed point, GFP), making the theory near-perturbative~\cite{Eichhorn:2018ydy}. Similarly to the case of non-abelian
gauge theories like QCD, the existence of an ultraviolet fixed point
is guaranteed whenever the ``paramagnetic interactions'' of the
action dominate over the diamagnetic ones~\cite{Nink:2012vd}. This
mechanism is at the basis of the (quantum) gravitational antiscreening.
The realization of the latter can be understood more intuitively from
the RG running of the Newton coupling. The structure of the beta function
of the dimensionless Newton coupling $g(k)=G(k)\,k^{2}$ in $d=4$
spacetime dimensions, $k$ being the RG scale, is
\begin{equation}
\beta_{g}=(2+\eta_{N})g\,,
\end{equation}
where $\eta_{N}=k\partial_{k}\log G(k)$ is the anomalous dimension
of the Newton coupling. The function $\eta_{N}$ depends on $g$ as
well as on all other (dimensionless) gravitational couplings. Further
note that $\eta_{N}$ depends on the RG scale $k$ only implicitly,
i.e., only through the RG running of the dimensionless gravitational
couplings. A necessary condition for the existence of a non-trivial
fixed point is that $\eta_{N}=-2$ for some values of the gravitational
couplings. Assuming that a non-trivial fixed point indeed exists in
the full (not truncated) theory space, the simple fact that $\eta_{N}<0$
at the non-trivial fixed point implies that the dimensionful Newton
coupling decreases with the RG scale $k$ and vanishes as $G(k)\sim g_{\ast}k^{-2}$
when $g(k)\to g_{\ast}$, where $g_{\ast}$ denotes the fixed-point
value of the dimensionless Newton coupling.

Although in the context of asymptotically safe gravity (ASG) the gravitational
antiscreening has a quantum origin and despite the different semantics
(Markov's asymptotic freedom of $G(k)$ vs asymptotic safety of $g(k)$),
this is the same principle advocated by Markov and Mukhanov in~\cite{Markov:1985py,Markov:1994hi}
as one possible way to soften or even remove the singularities affecting
the solutions of the Einstein equations. In the case of stellar black
holes the mechanism of singularity-avoidance is very intuitive: when
a collapsing star disappears behind its event horizon and its density
reaches planckian values, the gravitational interaction driving the
collapse becomes weaker. Therefore, under certain conditions, the
gravitational antiscreening can potentially halt the collapse and
prevent the formation of a spacetime singularity. Hints that the mechanism
of singularity-avoidance could be realized within the Asymptotic Safety
scenario for Quantum Gravity have been found in~\cite{Bonanno:1998ye,Bonanno:2000ep,Bonanno:2006eu,torres14,torres14b,Bonanno:2016dyv,Torres:2017ygl,Bonanno:2017kta,Bonanno:2017zen,Adeifeoba:2018ydh,Platania:2019kyx,Bonanno:2019ilz}
via the so-called RG-improvement procedure (see, e.g.,~\cite{Bonanno:2000ep,Bonanno:2012jy}
and references therein), in~\cite{Bosma:2019aiu} by means of non-perturbative
computations of quantum corrections to the Newtonian potential based
on the functional renormalization group (FRG) method, and in~\cite{Marunovic:2011zw,Marunovic:2012pr}
through a one-particle irreducible resummation of one-loop vacuum
fluctuations of non-minimally coupled, massless, scalar matter. In
analogy with the case of black-hole singularities, the gravitational
antiscreening could provide a solution to the problem of the initial
singularity~\cite{Kofinas:2015sna,Kofinas:2016lcz,Bonanno:2017gji}
in cosmology. Moreover, the existence of a regime where gravity is
approximately scale-invariant could be relevant in cosmology to provide
a natural explanation for the nearly-scale-invariant distribution
of temperature anisotropies in the Cosmic Microwave Background (CMB)
radiation~\cite{Bonanno:2001xi,Reuter:2005kb,Bonanno:2007wg,Weinberg:2009wa,Bonanno:2010mk,Bonanno:2015fga,Bonanno:2018gck}
(see also~\cite{Bonanno:2017pkg} for a recent review).

In this paper we review some of the main cosmological implications
of ASG based on the running of the gravitational
couplings. The rest of the present review is organized as follows.
Sect.~\ref{sec:3} summarizes the mechanism behind the renormalization group improvement and the scale-setting procedure. In Sect.~\ref{sec:5}
and~\ref{sec:6} we review the main implications of the gravitational
antiscreening in cosmology and inflation respectively. Finally, in
Sect.~\ref{sec:7} we summarize the state-of-the-start of phenomenology
of ASG, its main problems, and future perspectives.

\section{Running couplings and renormalization group improvement\label{sec:3}}

\subsection{Decoupling mechanism}

One of the strengths of ASG is the possibility of constructing a quantum
theory of gravity using the ``language'' of Quantum Field Theory
-- the standard framework to describe matter and all known fundamental
interactions within the Standard Model of particle physics. On the
one hand, this makes the connection between gravity and matter more
straightforward than in other approaches to quantum gravity and allows
to constrain the ultraviolet details of quantum gravity by verifying
systematically its consistency with low-energy experiments and observations
on the matter sector (see~\cite{Eichhorn:2017egq,Eichhorn:2018yfc}
for recent reviews). On the other hand, the computation of the quantum
corrections to the classical solutions of General Relativity requires
the knowledge of the gravitational quantum effective action $\Gamma_{0}^{grav}[g_{\mu\nu}]$.
The classical Einstein equations are replaced by the fully quantum
field equations
\begin{equation}
\frac{\delta\Gamma_{0}^{grav}[g_{\mu\nu}]}{\delta g_{\mu\nu}}=0\,.
\end{equation}
These are effectively classical field equations. Nonetheless, its
solutions $\langle g_{\mu\nu}\rangle$ actually incorporate all quantum
gravitational effects.

The computation of the effective action comes along with several technical
and conceptual issues. First, computing the effective action exactly
would require either to solve the gravitational path integral over
globally hyperbolic spacetimes or, equivalently, to solve the Functional
Renormalization Group (FRG) equation for the effective average action
$\Gamma_{k}$~\cite{Wetterich:1992yh,Morris:1993qb,ReuterWetterich}
\begin{equation}
k\partial_{k}\Gamma_{k}=\frac{1}{2}\mathrm{STr}\left\{ \left(\Gamma_{k}^{(2)}+\mathcal{R}_{k}\right)^{-1}k\partial_{k}\mathcal{R}_{k}\right\} \label{eq:floweq}
\end{equation}
and take the limit $k\to0$. The mathematical tools currently available
do not allow for an exact computation of the effective action, even
if important progress in this direction has been made in~\cite{Knorr:2018kog,Knorr:2019atm}.
Secondly, the effective action is a gauge- and parametrization-dependent
object: only physical observables will be independent of any gauge
choice, parametrization and regularization schemes. A third key issue,
related to the previous one, is that defining physical observables
in quantum gravity is still an outstanding open problem~\cite{Ambjorn:2005db,Ambjorn:2012jv}.

The \emph{decoupling mechanism}~\cite{Reuter:2003ca} might provide
a solution to some of the issues raised above. In what follows we
will summarize how this mechanism works, mostly following the arguments
and nomenclature in~\cite{Reuter:2003ca}. As is clear from Eq.~\eqref{eq:floweq},
the RG running of the effective action is determined by the modified
inverse propagator $\left(\Gamma_{k}^{(2)}+\mathcal{R}_{k}\right)$.
The regulator $\mathcal{R}_{k}\propto k^{2}$ is an effective mass-square
term that implements the Wilsonian shell-by-shell integration of fast
fluctuating modes: only fluctuations with momenta $p^{2}\gtrsim k^{2}$
are integrated out, resulting in the partially-quantized effective
action $\Gamma_{k}$. In the limit $k\to0$ all quantum fluctuations
are integrated out, so that $\Gamma_{0}$ coincides with the ordinary
effective action. At the basis of the decoupling mechanism is the
possibility that some infrared physical scales appearing in $\Gamma_{k}^{(2)}$,
such as a physical mass term, could compete and eventually overcome
the effect of the unphysical mass term in $\mathcal{R}_{k}$ in the
infrared. In this case there will be a threshold value of $k$ --
a decoupling scale $k_{dec}$ -- below which the running of $\Gamma_{k}$
is essentially frozen. The ``threshold effective action'' $\Gamma_{k_{dec}}$
and the ordinary effective action are thus expected to be approximately
the same. The identification of the infrared scale $k_{dec}$, if
any, could provide some of the (typically non-local) terms in $\Gamma_{0}$.
An emblematic example of this mechanism is massless scalar electrodynamics,
where $\Gamma_{k}^{(2)}=p^{2}+\lambda(k)\,\phi^{2}+\dots$ . The running
of the quartic coupling~$\lambda(k)$ is logarithmic, $\lambda(k)\sim\log k$
and therefore the decoupling should occur at $k_{dec}^{2}\sim\log(k_{dec})\,\phi^{2}$.
The effective action $\Gamma_{0}$ is thus expected to involve an
effective non-local interaction of the form $\phi^{4}\,\log(k_{dec}(\phi))$.
To leading order, this leads to the $\phi^{4}\,\log(\phi)$-interaction
appearing in the famous Coleman--Weinberg effective potential~\cite{cw}.
For other examples in QED and QCD see, e.g.,~\cite{1973migdal,1983adler,Dittrich:1985yb}
and references therein. When the effective action contains several
competing infrared scales (particle momenta, field strengths, spacetime
curvature, etc), identifying the threshold scale $k_{dec}$ becomes
a more involved task: $k_{dec}$ might be a complicated non-linear
function of all these scales or, in the best case, it might be given
by the one physical infrared scale which dominates over the others.
Conversely, a decoupling scale might not exist at all: this is the
case if there are no dominant infrared scales acting as an actual
physical cutoff. Therefore, the existence of $k_{dec}$ and its specific
form strongly depend on the physical system under consideration.

The procedure of identifying and replacing the RG scale $k$ with
a physical infrared scale, supposedly acting as a decoupling scale
$k_{dec}$, is known as \emph{RG improvement}. It aims at using the
RG running in order to incorporate leading-order quantum effects in
the dynamics of a classical system. In the example mentioned above,
the RG improvement is used \emph{at the level of the action} to obtain
the leading-order terms in the quantum effective action $\Gamma_{0}$.
Other forms of RG improvement are the RG improvement \emph{at the
level of the field equations} and \emph{at the level of the classical
solutions}; the latter allows, e.g., to derive the Uehling potential
from the RG-running of the electric charge~\cite{Dittrich:1985yb}. Even if the idea behind the decoupling mechanism seems to suggest that the RG improvement should be performed at the level of the action, this is typically considered as a sourch of ambiguity. \\
In the next subsection we will discuss the RG improvement in the
case of gravity and how to constrain the scale setting $k=k(x)$ based
on the symmetries of the theory.

We remark that the RG running of the gravitational couplings
extrapolated from FRG computations relies on the use of Euclidean
metrics. The implications of ASG obviously involve Lorentzian spacetimes.
It is thereby assumed that the scaling of the couplings and the existence
of an ultraviolet-attractive fixed point are not affected, at least
qualitatively, by the metric signature. Hints that this might indeed
be the case have been found in~\cite{Manrique:2011jc}.

\subsection{Renormalization group improvement and scale-setting procedure} \label{sect2b}

ASG relies on the existence of an ultraviolet-attractive fixed point
of the gravitational RG flow. In the Einstein-Hilbert truncation,
the scaling of the dimensionless Newton coupling $g(k)=G(k)\,k^{2}$
and cosmological constant $\lambda(k)=\Lambda(k)\,k^{-2}$ about the
NGFP $(g_{\ast},\lambda_{\ast})$ reads
\begin{equation}
\begin{cases}
g_{k}= & g_{\ast}+c_{1}e_{1}^{1}\left(\frac{k}{M_{Pl}}\right)^{-\theta_{1}}+c_{2}e_{2}^{1}\left(\frac{k}{M_{Pl}}\right)^{-\theta_{2}}\\
\lambda_{k}= & \lambda_{\ast}+c_{1}e_{1}^{2}\left(\frac{k}{M_{Pl}}\right)^{-\theta_{1}}+c_{2}e_{2}^{2}\left(\frac{k}{M_{Pl}}\right)^{-\theta_{2}}
\end{cases}\;,\label{eq:scalingUV}
\end{equation}
where $M_{Pl}\sim(8\pi G_{0})^{-1/2}$ is the reduced Planck mass,
the $c_{i}$ are integration constants labeling all possible RG trajectories,
$\textbf{e}_{i}$ are the eigenvectors of the stability matrix $\partial_{g_{i}}\beta_{j}|_{g_{\ast}}$
constructed using the beta functions $\beta_{j}$ of all dimensionless
couplings $g_{j}$, and $(-\theta_{i})$ are its eigenvalues. The
real part of the critical exponents $\theta_{i}$ determine the stability
properties of the NGFP. In the case of pure gravity, the critical
exponents $\theta_{1}$ and $\theta_{2}$ are typically a pair of
complex conjugate numbers with positive real part: this implies that
the NGFP is ultraviolet-attractive in the Einstein-Hilbert truncation.
In extended truncations, involving higher-derivative operators, it
has been shown that the NGFP comes with three relevant directions
associated with the volume, $R$, and 4th-order derivative operators
\cite{Codello:2006in,Codello:2007bd,Benedetti:2009rx,Benedetti:2009gn,Groh:2011vn,Falls:2014tra,Gies:2016con,Falls:2017lst,Falls:2018ylp,Alkofer:2018fxj}.
The number of relevant directions coincides with the number of free
parameters (the integration constants $c_{i}$) to be fixed by comparison
with observations, e.g., by requiring that in the infrared $8\pi G_{0}\sim M_{Pl}^{-2}$
and $\Lambda_{0}\sim3\cdot10^{-122}M_{Pl}^{2}$. Fixing the free parameters
in this way allows to select the ``RG trajectory realized by Nature''
\cite{Reuter:2004nx}. As it will be important in the applications
of ASG in cosmology, it worth mentioning that the values of the critical
exponents are influenced by the presence of matter: for instance,
in the Einstein-Hilbert truncation, the presence of minimally-coupled
free matter fields makes the critical exponents $\theta_{1,2}$ real
\cite{Dona:2013qba,Dona:2014pla,Biemans:2017zca,Christiansen:2017cxa,Alkofer:2018fxj}.

Neglecting the running of the matter couplings, the scale-dependent
Einstein-Hilbert action reads
\begin{equation}
S_{k}=\frac{1}{16\pi G_{k}}\int d^{4}x\sqrt{-g}\;(R-2\Lambda_{k})+S_{matter}\;.\label{eq:qaction}
\end{equation}
At some intermediate scale $k=k(x)$, the RG running modifies the
classical field equations by an effective energy-momentum tensor $\Delta t_{\mu\nu}\equiv G_{k}(\nabla_{\mu}\nabla_{\nu}-g_{\mu\nu}\square)G_{k}^{-1}$ which could encode, at an effective level, the vacuum polarization effects of the quantum gravitational
field~\cite{Reuter:2004nv},
\begin{equation}
G_{\mu\nu}=8\pi G_{k}T_{\mu\nu}-\Lambda_{k}g_{\mu\nu}+\Delta t_{\mu\nu}\;.\label{eq:modfieldeq}
\end{equation}
If there is no energy-momentum flow between the gravitational and
matter components of the theory, i.e., if the energy-momentum tensor
$T_{\mu\nu}$ is separately conserved, the cutoff function $k=k(x)$
is constrained by the modified (contracted) Bianchi identities~\cite{Reuter:2004nv,Reuter:2004nx,Babic:2004ev,Domazet:2012tw,Koch:2014joa}
\begin{align}
\nabla^{\mu}G_{\mu\nu} & =\left(8\pi G'_{k}T_{\mu\nu}-\Lambda'_{k}\,g_{\mu\nu}\right)\nabla^{\mu}k(x)+8\pi G_{k}\nabla^{\mu}T_{\mu\nu}+\nabla^{\mu}\Delta t_{\mu\nu}=0\;.\label{eq:bianchiID}
\end{align}
The above equation provides a ``consistency condition''~\cite{Reuter:2004nv,Reuter:2004nx,Babic:2004ev,Domazet:2012tw,Koch:2014joa}
which can be used, under certain assumptions and/or approximations,
to determine the scale-dependence $k=k(x)$. Specifically, we can
identify the following cases
\begin{itemize}
\item {CASE I}: \emph{RG improvement at the level of the field equations}

Performing the RG improvement at the level of the field equations
or solutions is equivalent to neglecting $\Delta t_{\mu\nu}$. Assuming
that the matter energy-momentum tensor is covariantly conserved, $\nabla_{\mu}T^{\mu\nu}=0$,
the shape of the function $k(x)$ is dictated by the condition
\begin{equation}
\left(8\pi G'_{k}T_{\mu\nu}-\Lambda'_{k}\,g_{\mu\nu}\right)\nabla^{\mu}k(x)=0\,.\label{eq:consistency1}
\end{equation}
The solution to this equation depends on the form of the energy-momentum
tensor. As it will be important in cosmology, let us focus on the
case of a perfect fluid with energy density~$\rho$ and pressure~$p$.
In this case $T_{\mu}^{\nu}=\mathrm{diag}(-\rho,p,p,p)$, with~$p=w\,\rho$,
and the above equation yields the condition
\begin{equation}
\frac{G'(k)}{G(k)}\left(\rho+\rho_{\Lambda}(k)\right)+\rho'_{\Lambda}(k)=0\,,
\end{equation}
with $\rho_{\Lambda}=\frac{\Lambda(k)}{8\pi G(k)}$. In the fixed
point regime, the scaling of the dimensionful Newton coupling and
cosmological constant reads
\begin{equation}
G(k)=g_{\ast}k^{-2}\;,\qquad\Lambda(k)=\lambda_{\ast}k^{2}\;\;,\label{eq:scaling}
\end{equation}
so that $k$ can be related to the matter energy density $\rho$~\cite{Babic:2004ev,Domazet:2010bk},
\begin{equation}
k^{4}=\left(\frac{8\pi g_{\ast}}{\lambda_{\ast}}\right)\,\rho\,.\label{eq:cutideq}
\end{equation}
In this setup the running gravitational couplings depend on the energy
density $\rho$ of the matter degrees of freedom: this is exactly
the \emph{ad hoc} assumption employed in~\cite{Markov:1985py,Markov:1994hi} to obtain a classical modification of General Relativity free of the problem of spacetime singularities.\\
The relation between $k$ and the energy density $\rho$ can also be understood in terms of the decoupling mechanism. If the spacetime is filled with a perfect fluid with energy density $\rho$, the gravitational action is complemented by a matter action $S_{matter}=-\int d^4x\sqrt{-g}\rho$. Due to the (minimal) coupling between the gravitational and matter degrees of freedom, the energy density $\rho$ enters the modified inverse propagator $(\Gamma_{k}^{grav,(2)}+\mathcal{R}_k)$ and could thus provide a decoupling scale $k_{dec}$ for the flow of the scale-dependent effective action $\Gamma_k^{grav}$. We note that the Ricci scalar~$R$ enters the modified inverse propagator as well and therefore it could also provide a decoupling scale. However, as we will see below, a scale setting $k^2\sim R$ satisfies the contracted Bianchi identities only if the RG improvement is performed at the level of the action.\\
In cosmology one can further limit the form of the effective metric to a Friedmann-Robertson-Walker (FRW) spacetime. Since $T_{\mu\nu}$ is assumed to be separately conserved, the energy density $\rho$ obeys the standard (i.e., classical) conservation equations and thus $\rho(a)=\rho_{0}(a(t)/a_{0})^{-3(1+w)}$ (note that the explicit time dependence of $\rho$ is determined by the form of the scale factor, and therefore it could differ from the classical case). In this case
\begin{equation}
k\propto a(t)^{-\frac{3}{4}(1+w)}\,.
\end{equation}
On the other hand, if $T_{\mu\nu}$ is not separately conserved, the
Bianchi identities
\begin{equation}
\left(8\pi G'_{k}T_{\mu\nu}-\Lambda'_{k}\,g_{\mu\nu}\right)\nabla^{\mu}k(x)+8\pi G_{k}\nabla^{\mu}T_{\mu\nu}=0
\end{equation}
do not add any additional constraint on the form of the cutoff function $k(x)$: the latter equation provides a generalized conservation equation allowing for an energy flow between the gravitational and matter degrees of freedom.

\item {CASE I}I: \emph{RG improvement at the level of the action}

If the RG improvement is performed at the level of the action, the
field equations contain an additional contribution encoded in the
gravitational energy-momentum tensor $\Delta t_{\mu\nu}$. Its variation reads~\cite{Koch:2014joa}
\begin{align}
\nabla^{\mu}\Delta t_{\mu\nu} & =G'_{k}(G_{k}^{-1}\Delta t_{\mu\nu})\nabla^{\mu}k(x)+G_{k}\underbrace{\nabla^{\mu}[(\nabla_{\mu}\nabla_{\nu}-g_{\mu\nu}\square)G_{k}^{-1}]}_{R_{\mu\nu}(G_{k}^{-1})'}\nabla^{\mu}k(x)\nonumber \\
 & =G'_{k}G_{k}^{-1}\left\{ (R_{\mu\nu}-\tfrac{1}{2}Rg_{\mu\nu}-8\pi G_{k}T_{\mu\nu}+\Lambda_{k}g_{\mu\nu})-R_{\mu\nu}\right\} \nabla^{\mu}k(x)\,.
\end{align}
The contracted Bianchi identities~\eqref{eq:bianchiID}, together
with the assumption that the energy-momentum tensor is separately
conserved, thus yield the condition
\begin{align}
\nabla^{\mu}G_{\mu\nu} & =\left(8\pi G'_{k}T_{\mu\nu}-\Lambda'_{k}\,g_{\mu\nu}\right)\nabla^{\mu}k(x)+8\pi G_{k}\underbrace{\nabla^{\mu}T_{\mu\nu}}_{0}\label{eq:bianchiID-1}\\
 & +G'_{k}G_{k}^{-1}\left\{ (R_{\mu\nu}-\tfrac{1}{2}Rg_{\mu\nu}-8\pi G_{k}T_{\mu\nu}+\Lambda_{k}g_{\mu\nu})-R_{\mu\nu}\right\} \nabla^{\mu}k(x)=0\;.\nonumber 
\end{align}
Note that in this case, due to the presence of $\Delta t_{\mu\nu}$,
the contribution from the energy-momentum tensor $T_{\mu\nu}$ cancels
out. A scaling relation of the form $k^{4}\sim\rho$~\cite{Babic:2004ev,Domazet:2010bk},
Eq.~\eqref{eq:cutideq}, is only valid under the assumption that $\Delta t_{\mu\nu}$
is negligible. We are thus left with the condition
\begin{equation}
\nabla^{\mu}G_{\mu\nu}=\left\{ G'_{k}G_{k}^{-1}(-\tfrac{1}{2}R+\Lambda_{k})-\Lambda'_{k}\right\} g_{\mu\nu}\,\nabla^{\mu}k(x)\equiv0\;.
\end{equation}
Diffeomorphism invariance thus requires~\cite{Reuter:2004nv,Domazet:2012tw,Koch:2014joa}
\begin{equation}
G'_{k}\,R=2(G'_{k}\Lambda_{k}-\Lambda'_{k}G_{k})\;.\label{eq:constraint}
\end{equation}
In the proximity to the NGFP, the couplings scale as in~\eqref{eq:scaling}
and the constraint~\eqref{eq:constraint} gives
\begin{equation}
k^{2}=\frac{R}{4\lambda_{\ast}}\;.\label{eq:cutid}
\end{equation}
This condition should also hold in the more general case of $f_{k}(R)$
theories, if the running of the gravitational couplings is approximated
with the corresponding fixed-point scaling~\cite{Domazet:2012tw,Platania:2019qvo}.\\
It is worth noting that the replacement $k^{2}\sim R$ in
the scale-dependent action~\eqref{eq:qaction} generates an effective
$f(R)$ action, whose analytical expression is determined by the running
of the gravitational couplings~\cite{Bonanno:2012jy,Domazet:2012tw,Hindmarsh:2012rc}. This fact is typically used to study effective inflationary models
in ASG, as it will be discussed in Sect.~\ref{sec:6}.
\end{itemize}

The RG improvement at the level of the field equations and at the level of the action lead to effective field equations which differ by a gravitational energy-momentum tensor~$\Delta t_{\mu\nu}$. Based on the idea behind the decoupling mechanism, performing the RG improvement at the level of the action would seem to be more natural. However, the possibility of choosing between different forms of RG improvement is considered as a source of ambiguity. The identification of the decoupling scale $k_{dec}$ is a second possible source of ambiguity: in the case of gravity, there are multiple scales that could potentially act as a decoupling scale for the flow of the scale-dependent gravitational effective action. However, if the matter energy-momentum tensor is covariantly conserved, the form of the cutoff function $k(x)$ is constrained by the contracted Bianchi identities. 

Phenomenological implications of ASG have been explored in the literature by means of RG-improved cosmological and astrophysical models. Although these models are not expected to provide precise and quantitative predictions, they are expected to capture qualitative features of the modifications of Einstein gravity induced by ASG. 

Keeping strenghts and limitations of the RG-improvement procedure in mind, in the next sections we will review some of the main phenomenological implications of ASG based on models of RG-improved cosmology.

\section{RG-improved cosmologies\label{sec:5}}

It is an old idea that the gravitational couplings could depend on
the cosmic time and that this time-dependence could have implications
in cosmology~\cite{Reuter:1986wm}. In the context of RG-improved
cosmologies this time-dependence arises from the RG running of the
gravitational couplings. 

In this section we will review some of the
main cosmological implications of ASG~\cite{Bonanno:2001xi,Bonanno:2002zb,Reuter:2005kb,Bonanno:2007wg,Bonanno:2017gji}.
Each subsection focuses on the cosmological implications obtained
in different regimes and/or under different assumptions.

\subsection{Early-universe cosmology: the NGFP regime\label{subsec:Early-universe}}

We first focus on the RG-improved cosmological dynamics in the NGFP
regime, following the analysis in~\cite{Bonanno:2001xi}. The starting
point is the assumption that the effective metric, solution to the
fully quantum equation of motion, is a homogeneous and isotropic FRW
universe,
\begin{equation}
ds^{2}=-dt^{2}+a^{2}(t)\left[\frac{dr^{2}}{1-K\,r^{2}}+r^{2}d\Omega^{2}\right]\,.
\end{equation}
The matter degrees of freedom are encoded in a perfect fluid with
energy-momentum tensor $T_{\mu}^{\nu}=\mathrm{diag}(-\rho,p,p,p)$
and equation of state $p=w\rho$. In this setup quantum-gravitational
fluctuations can only modify the effective dynamics of the scale factor
$a(t)$. Performing an RG improvement at the level of the fields equations
(in the Einstein-Hilbert truncation) yields the modified Friedmann
equation 
\begin{equation}
\left(\frac{\dot{a}}{a}\right)^{2}+\frac{K}{a^{2}}=\frac{\Lambda[k(x)]}{3}+\frac{8\pi G[k(x)]}{3}\rho\,.
\end{equation}
In this subsection we focus on the case where the energy-momentum
tensor is separately conserved, as originally assumed in~\cite{Bonanno:2001xi}.
In this case no flow of energy between the gravitational and matter
sector is possible (the case in which $T_{\mu\nu}$ is not covariantly
conserved has been studied in~\cite{Bonanno:2007wg} and will be discussed
in Sect.~\ref{subsec:Entropy-production}). The conditions $\nabla_{\mu}G^{\mu\nu}=0$
and $\nabla_{\mu}T^{\mu\nu}=0$ thus result in the ``standard''
conservation equation, $\dot{\rho}=-3H\rho(1+w)$, and in the consistency
condition~\eqref{eq:consistency1}. The latter can be split in the
following equations
\begin{equation}
\partial_{t}\Lambda+8\pi\rho\,\partial_{t}G=0\,,\label{eq:consistency}
\end{equation}
\begin{equation}
h^{\mu\nu}(\Lambda_{;\nu}-8\pi p\,G_{;\nu})=0\,,
\end{equation}
where $h_{\mu\nu}=g_{\mu\nu}+u_{\mu}u_{\nu}$ is the projection tensor
onto the tangent 3-space orthogonal to the 4-velocity $u^{\mu}$ of
an observer comoving with the cosmological fluid. Provided that $G$
and $\Lambda$ do not vary along the hypersurfaces orthogonal to $u^{\mu}$,
i.e., $k=k(t)$, the latter relation is identically satisfied. The
cosmological evolution of the universe is then (over-) determined
by the system of equations~\cite{Bonanno:2001xi}
\begin{equation}
\begin{split} & \frac{\dot{a}^{2}}{a^{2}}+\frac{K}{a^{2}}=\frac{\Lambda(t)}{3}+\frac{8\pi G(t)}{3}\rho\,,\\
 & \dot{\rho}=-3H\rho(1+w)\,,\\
 & \dot{\Lambda}(t)+8\pi\rho\,\dot{G}(t)=0\,.
\end{split}
\label{eq:cosmosystem}
\end{equation}
The consistency condition~\eqref{eq:consistency} can thus be used
to constrain the form of the cutoff function $k(t)$. Assuming that
in the early universe the growth of the scale factor follows a power
law, Eq.~\eqref{eq:consistency} is consistent with a scale dependence
of the form $k=\xi_{t}\,t^{-1}$, where $\xi_{t}$ is a positive constant
\cite{Bonanno:2001xi}. In particular, the consistency condition~\eqref{eq:consistency}
fixes the value of the free parameter $\xi_{t}$ in terms of fixed-point
quantities. As it will become clear soon, the scale-setting $k\sim t^{-1}$
employed in~\cite{Bonanno:2001xi} is also compatible with the one
in Eq.~\eqref{eq:cutideq}, that was derived and discussed in subsequent
studies~\cite{Babic:2004ev,Domazet:2010bk}.

In what follows we will only consider the case of a spatially flat
universe, $K=0$. Approximating $G(k)$ and $\Lambda(k)$ with their
fixed-point scaling, Eq.~\eqref{eq:scaling}, with~$k(t)=\xi_t t^{-1}$, the cosmological system~\eqref{eq:cosmosystem} can be solved analytically. The modified Friedmann and conservation equations admit a family of solutions where the scale factor $a(t)$ and the density $\rho(t)$ scale as power laws. Their dependence on the constant $\xi_t$ can then be eliminated by imposing the consistency condition~\eqref{eq:consistency}, which leads to the relation
\begin{equation}
\xi_{t}=\sqrt{\frac{8}{3(1+w)^{2}\lambda_{\ast}}}\,.
\end{equation}
Using this expression for $\xi_t$, the family of cosmological solutions associated with the RG-improved system~\eqref{eq:cosmosystem} reads~\cite{Bonanno:2001xi}
\begin{equation}
\begin{split} & a(t)=\left(\frac{9}{64}(1+w)^{4}g_{\ast}\lambda_{\ast}M\right)^{\frac{1}{3(1+w)}}t^{\frac{4}{3(1+w)}}\,,\\
 & \rho(t)=\frac{8}{9\pi(1+w)^{4}g_{\ast}\lambda_{\ast}}\,t^{-4}\,,\\
 & G(t)=\left(\frac{3}{8}(1+w)^{2}g_{\ast}\lambda_{\ast}\right)\,t^{2}\,,\\
 & \Lambda(t)=\left(\frac{3}{8}(1+w)^{2}\right)^{-1}\,t^{-2}\,,
\end{split}
\label{eq:cosmosol}
\end{equation}
where $M$ is an integration constant. These solutions depend on fixed-point
quantities through the combination $(\lambda_{\ast}g_{\ast})$, which
is known to be scheme independent~\cite{Dou:1997fg,Lauscher:2001ya}. Moreover,
the solutions do not depend on the infrared values of the gravitational
couplings, reflecting the universal behavior of the RG flow at the
NGFP. Note that, provided that the scale factor follows a power law
scaling (i.e., $w\neq-1$), the cutoff identification $k=\xi_{t}t^{-1}$
is equivalent to the scale settings $k=\xi_{h}H(t)$ (even at a classical
level), $k=\xi_{\rho}\rho^{1/4}$ (due to the modified power law of
the scale factor, and in accordance with the discussion in Sect.~\ref{sec:3})
and, in a RG-improved radiation-dominated era, to $k\sim\xi_{a}a^{-1}$,
with
\begin{equation}
\xi_{t}^{2}=\frac{8}{3(1+w)^{2}\lambda_{\ast}}\,,\qquad\xi_{h}^{2}=\frac{3}{2\lambda_{\ast}}\,,\qquad\xi_{\rho}^{2}=\frac{8\pi g_{\ast}}{\lambda_{\ast}}\,,\qquad\xi_{a}^{2}=\frac{g_{\ast}M}{\lambda_{\ast}}\,.
\end{equation}
This magic can only occur in the proximity of a critical fixed point,
where physical quantities should vary as power laws of a unique scale.
Away from the NGFP, the complete solution to the cosmological system~\eqref{eq:cosmosystem} can only be obtained numerically. This has
been done in~\cite{Reuter:2005kb}. In this case the cutoff function
$k(t)$ is obtained by solving the full beta functions for the gravitational
couplings and the consistency condition~\eqref{eq:consistency} numerically.
In particular, in~\cite{Reuter:2005kb} it has been shown that the dynamical
cutoff $k(t)$ is well approximated by the Hubble constant, $k\sim H(t)$,
for any value of the cosmic time~$t$.

The fixed-point scaling of gravitational couplings modifies the power-law
scaling of the scale factor, Eq.~\eqref{eq:cosmosol}, so that also
the causal structure of the spacetime is modified at early times.
At it can be easily seen, provided that $w\leq1/3$, there is no particle
horizon~\cite{Bonanno:2001xi}: quantum effects enlarge the extension
of the light-cones such that events occurring at the decoupling era
are causally influenced by all points belonging to the hypersurface
$t=0$~\cite{Bonanno:2001xi}.  Nonetheless, if $w=1/3$ the deceleration
parameter is zero, i.e., no inflation occurs in a RG-improved radiation-dominated
epoch. This problem can be overcome by relaxing the assumptions and/or
improving the approximations made in~\cite{Bonanno:2001xi}: a period
of inflation can occur in RG-improved cosmologies if the gravitational
and matter degrees of freedom can exchange energy~\cite{Bonanno:2007wg}
(this will be discussed in Sect.~\ref{subsec:Entropy-production})
and/or when the gravitational effective energy-momentum tensor $\Delta t_{\mu\nu}$
discussed in Sect.~\ref{sec:3} is taken into account, i.e., when
the RG improvement is performed at the level of the action~\cite{Bonanno:2010bt,Hindmarsh:2012rc,Bonanno:2015fga,Bonanno:2018gck}
(this point will be discussed in detail in Sect.~\ref{sec:6}).

The gravitational effective energy-momentum tensor $\Delta t_{\mu\nu}$
is also crucial to make the cosmological evolution non-singular: in
the setting introduced above, the RG-improved cosmological evolution~\eqref{eq:cosmosol} is still singular, as scale factor vanishes at
$t=0$. The $\Delta t_{\mu\nu}$-term in the effective field equations~\eqref{eq:modfieldeq} might mimic the effect of higher-order operators
in the gravitational effective action~\cite{Bonanno:2010bt,Hindmarsh:2012rc,Bonanno:2015fga,Bonanno:2018gck}
and, as discussed in~\cite{Lehners:2019ibe}, higher-order operators
could be crucial to explain the early-universe evolution and its initial
conditions. As it will be discussed in Sect.~\ref{subsec:bounce},
starting from the Einstein-Hilbert truncation and performing the RG improvement
at the level of the action gives rise to additional terms in the modified
Friedmann equations (corresponding to the additional $\Delta t_{\mu\nu}$-term
in the effective field equations~\eqref{eq:modfieldeq}) which allow
for a non-singular cosmological evolution for any value of the spatial
curvature $K$~\cite{Bonanno:2017gji}.

\subsection{Entropy production\label{subsec:Entropy-production}}

The results reviewed in the previous subsection are based on the assumption
that the matter energy-momentum tensor $T_{\mu}^{\nu}=\mathrm{diag}(-\rho,p,p,p,)$
is covariantly conserved. In a series of works~\cite{Bonanno:2007wg,Bonanno:2008xp,Bonanno:2010mk},
the possibility of an energy flow between the gravitational and matter
sectors have been considered and its implications have been explored
in detail. The main result is that, under certain assumptions, this
energy flow could provide an explanation for the production of entropy
during the primordial evolution of the universe.

If the matter energy-momentum tensor $T_{\mu}^{\nu}=\mathrm{diag}(-\rho,p,p,p,)$
is not separately conserved, the continuity equation arising from
the contracted Bianchi identity reads
\begin{equation}
\dot{\rho}+3H(\rho+p)=-\frac{\dot{\Lambda}+8\pi\rho\dot{G}}{8\pi G}
\end{equation}
and can be written in the form
\begin{equation}
\frac{\mathrm{d}}{\mathrm{d}t}(\rho a^{3})+p\,\frac{\mathrm{d}}{\mathrm{d}t}(a^{3})=\mathcal{P}(t)\equiv-\frac{\dot{\Lambda}+8\pi\rho\dot{G}}{8\pi G}a^{3}\,.
\end{equation}
Identifying $U=\rho a^{3}$ as the energy encapsulated in the proper
volume $V=a^{3}$, the latter equation assumes the form of the first
law of thermodynamics
\begin{equation}
\mathrm{d}U+p\,\mathrm{d}V=\mathcal{P}(t)\mathrm{d}t\equiv T\mathrm{d}S\,.
\end{equation}
In classical cosmology $\mathcal{P}(t)=0$ and the evolution of the
universe is regarded as an adiabatic process. The introduction of
time-varying gravitational couplings entails instead a variation of
the entropy generated by the continuous energy flow between the gravitational
and matter sectors,
\begin{equation}
\frac{\mathrm{d}S}{\mathrm{d}t}=T^{-1}\mathcal{P}(t)\,.
\end{equation}
Specifically, the production of entropy during the expansion of the
universe requires $(\dot{\Lambda}+8\pi\rho\dot{G})\leq0$. As $k(t)$
decreases with the cosmic time $t$, it follows that $\dot{\Lambda}\leq0$
and $\dot{G}\geq0$. Entropy production thus requires the speed of
variation of the running cosmological constant to overcome that of
the Newton coupling, such that $-\dot{\Lambda}/\dot{G}\geq8\pi\rho$.

Assuming that the universe is initially dominated by radiation and
that the primordial evolution is an approximately adiabatic process,
$\mathcal{P}(t)\approx0$, the standard equilibrium conditions relating
the thermodynamic variables $(\rho,V,T)$ can still be used. In this
case $\rho(T)\propto T^{4}$ and therefore $S(t)\propto a^{3}\rho^{3/4}+\mathrm{const.}$.
The precise behavior of $S(t)$ can be obtained by solving the RG-improved
Friedmann and continuity equations. This requires to set the scaling
relation $k=k(t)$. Since the cosmological evolution is assumed to
be approximately adiabatic, the consistency condition~\eqref{eq:consistency} should be approximately verified and,
based on the results of~\cite{Reuter:2005kb}, a cutoff function of
the form $k(t)=\xi_{h}\,H(t)$ could still be employed.

Focusing on the ``NGFP era'', where $G=g_{*}k^{-2}$ and $\Lambda=\lambda_{*}k^{2}$,
the scale setting $k(t)\sim\xi_{h}\,H(t)$ leads to an effective cosmological
evolution where the scale factor varies as a power law $a(t)\propto t^{\alpha}$,
with 
\begin{equation}
\alpha=\frac{2}{(3+3w)(1-\Omega_{\Lambda}^{*})}
\end{equation}
and $\Omega_{\Lambda}^{*}=\xi_{h}^{2}\lambda_{\ast}/3$. The corresponding
deceleration parameter is $q=\alpha^{-1}-1$. Imposing the consistency
condition~\eqref{eq:consistency} fixes $\Omega_{\Lambda}^{*}=1/2$
and gives back the solution obtained in~\cite{Bonanno:2001xi} and
discussed in Sect.~\ref{subsec:Early-universe}: allowing for an energy
flow between the gravitational and matter degrees of freedom, removes
this additional constraint and leads to a family of cosmological solutions,
each characterized by a fixed value of $\Omega_{\Lambda}^{*}$. However,
since $\mathcal{P}(t)\approx0$ is assumed, the value of $\Omega_{\Lambda}^{*}$
should not differ to much from that obtained by imposing the consistency
condition, i.e., $\Omega_{\Lambda}^{*}\sim1/2$. As the value of $\lambda_{\ast}$
derived from FRG computations is of order $\mathcal{O}(1)$, this
also implies that $\xi_{h}=\mathcal{O}(1)$. The transition to the
classical FRW cosmology thus occurs when $k(t_{\mathrm{tr}})=\xi_{h}H(t_{\mathrm{tr}})\sim M_{Pl}$
which, using the fact that $\xi_{h}\sim1$, implies that the parameter
$\alpha$ sets the ratio between the transition time $t_{\mathrm{tr}}$
and the Planck time, $t_{\mathrm{tr}}=\alpha\,t_{Pl}$. We thus learn
that if $\alpha>1$ the transition to the classical regime occurs
before the Planck time, $t_{\mathrm{tr}}>t_{Pl}$.

In a RG-improved radiation-dominated epoch, the production of entropy
is given by the power law $\mathcal{P}(t)\propto(\alpha-1)t^{3\alpha-4}$,
so that 
\begin{equation}
S(t)\propto(t^{3(\alpha-1)}+\mathrm{const.})\,.
\end{equation}
In particular, the condition $dS\geq0$ is met if $\alpha\geq1$.
The case $\alpha=1$ ($\Omega_{\Lambda}^{*}=1/2$) describes a universe
where the gravitational and matter degrees of freedom are decoupled
or, equivalently, $\mathcal{P}(t)=0$. If instead $\alpha>1$, the
variation of $\Lambda$ dominates over the variation of $G$, resulting
in a net entropy production during the Planck era (NGFP regime). Specifically, assuming $S(0)=0$, within this model the entropy production can be entirely explained by the variation of $\Lambda$.

We highlight that the condition $\alpha>1$, necessary to generate
entropy in the early-universe expansion, is the same condition needed
in order to produce a period of power-law inflation. In a radiation-dominated
epoch this is condition is satisfied for $\Omega_{\Lambda}^{*}>1/2$.
If $\alpha=1$ ($\Omega_{\Lambda}^{*}=1/2$, case analyzed in~\cite{Bonanno:2001xi})
no inflation occurs but, as in the case $\alpha>1$, no particle horizon
exists. Therefore, within this simplified model , if $\Omega_{\Lambda}^{*}\gtrsim1/2$
(or, equivalently, $\alpha\gtrsim1$) the RG running of the gravitational
couplings during the Planck epoch can explain the production of entropy
and, at the same time, provide a period of power-law inflation.

\subsection{Cosmological singularities and bouncing cosmologies in ASG \label{subsec:bounce}}

The existence of an ultraviolet-attractive NGFP entails a weakening
of the gravitational interaction at high energies. It is then natural
to ask whether this weakening can lead to non-singular cosmologies.
While a definite answer in the context of ASG is still out of reach,
the mechanism underlying a possible singularity resolution might be
captured by a simple model embedding the running of the gravitational
couplings in the spacetime dynamics~\cite{Bonanno:2017gji}.

Following the discussion in Sect.~\ref{sec:3}, the running of the
gravitational couplings in the Einstein-Hilbert action generates an
additional term $\Delta t_{\mu\nu}$ in the modified field equations.
As we have seen, neglecting this term and introducing the running
couplings at the level of the field equations leads to a family of
RG-improved cosmologies admitting a period of power-law inflation
and explaining the entropy production in terms of the energy flow
between the gravitational and matter sectors~\cite{Bonanno:2007wg}.
These cosmologies are however singular. As shown explicitly in~\cite{Bonanno:2017gji}, taking into account the correction $\Delta t_{\mu\nu}$ to the effective
field equations~\eqref{eq:modfieldeq} might modify this conclusion.

It is assumed that the universe is homogeneous and isotropic and that
the energy-momentum tensor is covariantly conserved. Introducing the
running of the gravitational couplings at the level of the (Einstein-Hilbert) action yields the modified Friedmann equation~\cite{Bonanno:2017gji}
\begin{equation}
\frac{\dot{a}^{2}}{a^{2}}+\frac{K}{a^{2}}=\left(\frac{\Lambda}{3}+\frac{8\pi G}{3}\rho\right)-\eta_{N}\frac{\dot{a}^{2}}{a^{2}}\,,\label{eq:modfri2}
\end{equation}
where $\eta_{N}=-\frac{\partial\log G(a)}{\partial\log a}$. The key
difference between this model~\cite{Bonanno:2017gji} and the one
analyzed in the previous subsections~\cite{Bonanno:2001xi,Reuter:2005kb}
lies in the presence of an additional term, $\Delta=-\eta_{N}H^{2}$,
in the modified Friedmann equation. Since $\eta_{N}\to0$ as the RG
flow approaches the perturbative regime ($a(t)\gg L_{Pl}$), $\Delta$
vanishes in this limit. Going back in time, $\eta_{G}$ varies from
its classical value $\eta_{N}=0$ to the fixed-point value $\eta_{N}=-2$
(reached when $k\to\infty$). 

In~\cite{Bonanno:2017gji} a scaling $k^{2}\propto R\sim a^{-2}$ was assumed. This approximation is valid
if the scale factor undergoes a period of exponential growth at early
times -- an assumption that can be verified \emph{a posteriori}. 

The analytical form of Newton coupling $G(k)$ and cosmological constant $\Lambda(k)$ can be obtained by solving the corresponding flow equations. Using the beta functions derived from the Wetterich equation~\eqref{eq:floweq}, it has been shown~\cite{Bonanno:2000ep} that they admit the following approximate solutions 
\begin{equation}\begin{aligned} &G(k) \simeq G_{0}\left(1+G_{0} g_{*}^{-1} k^{2}\right)^{-1},\\ &\Lambda(k) \simeq \Lambda_{0}+\lambda_{*} k^{2}, \end{aligned}\label{apprela}\end{equation}
where $(G_{0},\Lambda_{0})$ are the low-energy values of the gravitational couplings. Replacing the relations~\eqref{apprela},
with $k\sim a^{-2}$, in Eq.~\eqref{eq:modfri2} and assuming that the universe is initially dominated by radiation, it can be easily seen~\cite{Bonanno:2017gji} that the field equations~\eqref{eq:modfri2} admit non-singular cosmological
solutions with minimum radius
\begin{equation}a_{b}^{2}=-\frac{G_{0} \Lambda_{0}+g_{*}\left(\lambda_{*}-3 K\right)}{2 g_{*} \Lambda_{0}} \pm \sqrt{\left(\frac{G_{0} \Lambda_{0}-g_{*}\left(\lambda_{*}-3 K\right)}{2 g_{*} \Lambda_{0}}\right)^{2}-\frac{8 \pi M G_{0}}{\Lambda_{0}}},\end{equation}
where $M$ is an integration constant. Depending on the values of
the fixed-point parameters and on the spatial curvature $K$, both
a bouncing cosmology or an emergent universe scenario could in principle be realized~\cite{Bonanno:2017gji}. This happens if
\begin{equation} M\leq \frac{\Lambda_{0}}{8 \pi G_{0}} \left(\frac{G_{0} \Lambda_{0}-g_{*}\left(\lambda_{*}-3 K\right)}{2 g_{*} \Lambda_{0}}\right)^{2} 
\end{equation}
and $a_b^2>0$. In this case the universe undergoes a period of inflation at early times, where the scale factor grows exponentially~\cite{Novello:2008ra}.
Otherwise, if $a_{b}$ is not real, the universe is singular and a
period of exponential growth of the scale factor is not possible (unless
other degrees of freedom are introduced). However this would invalidate the initial assumption that $k^{2}\propto R\sim a^{-2}$ and a separate analysis would be required. This model thus shows how the gravitational antiscreening, encoded in the RG running of the gravitational couplings and in the presence of additional terms in the effective Friedmann equation, could lead to non-singular cosmologies and a period of exponential growth of the universe at early times.

\section{Inflation in Asymptotically Safe Gravity\label{sec:6}}

\subsection{The idea behind ``Asymptotically Safe Inflation''\label{subsec:asinfl}}

Primordial quantum fluctuations occurring in the pre-inflationary
epoch have left indelible imprints, which we measure today in the
form of tiny temperature anisotropies, $\delta T/T\sim10^{-5}$, in
the CMB radiation: according to the standard cosmological model, the
inhomogeneities in the CMB can be traced back to the primordial quantum
fluctuations in the pre-inflationary era. These fluctuations were
subsequently amplified and smoothed out by the exponential growth
of the universe, thus resulting in small density fluctuations at the
last scattering surface. The distribution of temperature anisotropies
in the sky could thus give us indirect information on the physics
of the very early universe.

In momentum space, the power spectra of scalar and tensorial perturbations
are written as follows
\begin{equation}
\mathcal{P}_{s}(k)\simeq A_{s}\left(\frac{k}{k_{\ast}}\right)^{n_{s}-1}\;,\qquad\mathcal{P}_{t}(k)\simeq A_{t}\left(\frac{k}{k_{\ast}}\right)^{n_{t}}\;,\label{eq:Sps}
\end{equation}
where $k=|\mathbf{k}|$ is the norm of the 3-momentum $\mathbf{k}$
and $k_{\ast}\sim0.05\text{Mpc}^{-1}$ is a reference scale. The spectral
index $n_{s}$ and the tensor-to-scalar ratio $r\equiv A_{t}/A_{s}$
can be obtained from observational data. In particular, the most recent
observations to date~\cite{Akrami:2018odb} constrain the spectral
index to be~$n_{s}=0.9649\pm0.0042$ at $68\%$ confidence level,
and limit the tensor-to-scalar ratio to values~$r<0.064$. Note that
although the scalar power spectrum is almost scale invariant, 
perfect scale invariance (corresponding to~$n_{s}=1$) is excluded.\\

The existence of a NGFP in the RG flow of gravity could provide a
natural and intuitive explanation for the nearly-scale invariance
of the power spectrum of temperature fluctuations in the CMB. Close
to the NGFP, the effective background graviton propagator behaves
as $G(p)\sim1/p^{d-2-\eta_{N}}$~\cite{Lauscher:2005qz}. In $d=4$,
the asymptotic-safety condition requires the anomalous dimension of
the Newton coupling to approach the value $\eta_{N}=-2$ in the ultraviolet
limit. In this case the background graviton propagator in coordinate
space scales as $\mathcal{G}(x,y)\sim\log|x-y|^{2}$ at the NGFP~\cite{Lauscher:2005qz}.
Assuming that the temperature fluctuations are entirely due to the
amplification of the quantum fluctuation of the spacetime geometry
during inflation and that these fluctuations are generated during
the Planck era, the corresponding density fluctuations $\delta\rho$
are characterized by a two-point correlation function~\cite{Lauscher:2002sq,Lauscher:2005qz,Bonanno:2007wg}
\begin{equation}
\xi(\mathbf{x})=\langle\delta\hat{\rho}(\mathbf{x}+\mathbf{y})\delta\hat{\rho}(\mathbf{y})\rangle\propto\langle\delta\mathbf{R}(\mathbf{x}+\mathbf{y},t)\delta\mathbf{R}(\mathbf{y},t)\rangle\sim|\mathbf{x}|^{-4},
\end{equation}
where $\delta\hat{\rho}=\delta\rho/\bar{\rho}$ is the fractional
density fluctuation field and $\delta\mathbf{R}(\mathbf{y},t)$ stands
for the fluctuation of the scalar curvature (or any component of the
Riemann or Einstein tensor~\cite{Bonanno:2001xi}), induced by a metric
fluctuation. The power spectrum in momentum space is given by the
3-dimensional Fourier transform
\begin{equation}
|\delta_{k}|^{2}=V\int d^{3}x\,\xi(\mathbf{x})e^{-i\mathbf{k}\cdot\mathbf{x}}.
\end{equation}
The spectral index $n_{s}$ defines the power-law scaling of the power
spectrum, $|\delta_{k}|^{2}\propto|\mathbf{k}|^{n_{s}}$. Thus the
scaling $\xi(\mathbf{x})\sim|\mathbf{x}|^{-4}$ gives rise to a perfectly
scale invariant power spectrum, with $n_{s}=1$~\cite{Lauscher:2002sq,Bonanno:2002zb,Lauscher:2005qz,Bonanno:2007wg}.
The exact scale invariance of the power spectrum reflects the exact
scale invariance of the theory at the NGFP. It is thereby possible
that the nearly-scale-invariance of the scalar power spectrum $\mathcal{P}_{s}(k)$
is due to the nearly-scale invariant flow of RG trajectories in the
proximity of the NGFP.
This observation~\cite{Lauscher:2002sq,Bonanno:2002zb,Lauscher:2005qz,Bonanno:2007wg}
was the starting point for a number of studies looking for the existence
of (unstable) deSitter solutions in ASG~\cite{Weinberg:2009wa,Bonanno:2010bt,Falls:2016wsa},
giving rise to a sufficiently long period of ``NGFP-driven inflation''
\cite{Bonanno:2007wg}. The inflationary scenario arising from this
mechanism is sometimes called ``Asymptotically Safe Inflation''.

\subsection{Starobinsky model and RG-running in quadratic gravity}

Among all proposed inflationary models~\cite{inflationaris}, the
Starobinsky model is certainly one of the most appealing: it is a
zero-parameters model and is compatible with the current observational
data~\cite{Akrami:2018odb}. The Starobinsky model relies on the inclusion
of an $R^{2}$-term in the gravitational action. This is the minimal
modification of Einstein gravity needed to produce inflation. From
the point of view of ASG, focusing on an $f(R)$-truncation, the quadratic
gravity Lagrangian
\begin{equation}
\mathcal{L}=\frac{1}{16\pi G}(R-2\Lambda-B\,R^{2})\label{eq:quadgrav}
\end{equation}
should comprise all relevant couplings of the theory (with respect
to the NGFP): according to the studies of the renormalization group
flow of $f(R)$ theories (in pure gravity), the NGFP comes with three
relevant directions -- those associated with the couplings $(G,\Lambda,B)$
\cite{Falls:2014tra,Falls:2017lst,Falls:2018ylp,Alkofer:2018fxj}.
The latter are the only free parameters of the theory: every RG trajectory
is uniquely identified by the infrared values of the scale-dependent
couplings $(G(k),\Lambda(k),B(k))$. 
A key question is whether there
exists an RG trajectory matching the infrared values of these couplings.
Moreover, it is interesting to understand whether Starobinsky inflation
can be realized naturally in the context of ASG. For this to happen
the sign of the coupling $B$ is crucial, as the Starobinsky model
requires $B$ to be negative. In addition, it is a key requirement
that classical Einstein gravity is recovered at low energies. Studying
the RG flow of the couplings $(G(k),\Lambda(k),B(k))$ in the quadratic
truncation~\eqref{eq:quadgrav}, it has been shown~\cite{Gubitosi:2018gsl}
that there exists an RG trajectory such that the observational constraints
\begin{align}
 & B(k_{\mathrm{infl}}\sim H_{\mathrm{infl}}=10^{22}eV)=-1.7\cdot10^{-46}eV^{-2}\,,\nonumber \\
 & G(k_{\mathrm{lab}}\sim10^{-5}eV)=6.7\cdot10^{-57}eV^{-2}\,,\\
 & \Lambda(k_{\mathrm{Hubble}}\sim H_{0}=10^{-33}eV)=4\cdot10^{-66}eV^{2}\,,\nonumber 
\end{align}
are all fulfilled. We refer the reader to the original paper~\cite{Gubitosi:2018gsl}
for the details of the computation. The coupling $B(k)$ is initially
($k\to\infty$) positive, but it turns negative along the RG flow:
the transition scale is $k\sim10^{23}GeV$ (well above the Planck
scale), so that at inflationary scales the action~\eqref{eq:quadgrav}
matches that of the Starobinsky model~\cite{Gubitosi:2018gsl}. Below
the Planck scale, the couplings $B(k)$ and $G(k)$ vary by many orders
of magnitude in a very short RG-time, $t=\log k$. Their observed
constant values are thus reached at inflationary scales. The cosmological
constant instead keeps running even after the end of inflation: at
inflationary scales its magnitude is $\sim4\cdot10^{30}eV^{2}$, while
its observed constant value is only reached at $k\sim10^{-2}eV$~\cite{Dou:1997fg,Gubitosi:2018gsl}.

\subsection{Constraints from Planck data in gravity-matter systems}

In this subsection we review the results in~\cite{Bonanno:2018gck,Platania:2019qvo}.
As the initial conditions for inflation are placed at trans-Planckian
scales and since the effective action at inflationary scales depends
on how the RG-trajectory realized by Nature emerges from the NGFP,
the Planck data on CMB anisotropies can in principle put constraints
on the universality properties of the gravitational RG flow. The latter
are encoded in the critical exponents $\theta_{i}$ governing the
scaling of the gravitational couplings in the vicinity of the NGFP.
In turn, the specific values of the critical exponents depend on the
number of scalar, Dirac and vectors fields in the theory~\cite{Dona:2013qba,Biemans:2017zca,Alkofer:2018fxj}.
The observational constraints on the spectral index $n_{s}$ and tensor-to-scalar
ratio $r$ could then be used to put constraints on the primordial
matter content of the universe. Introducing the running of the gravitational
couplings at the level of the (Einstein-Hilbert) action~\cite{Bonanno:2012jy,Hindmarsh:2012rc}
provides a simple toy model to understand whether and how this mechanism
is realized~\cite{Bonanno:2018gck,Platania:2019qvo}.

We restrict ourselves to the Einstein-Hilbert truncation, where the
scaling of the gravitational couplings about the NGFP is that given
in Eq.~\eqref{eq:scalingUV}. Following the discussion in Sect.~\ref{sec:3}, close to the NGFP the consistency condition~\eqref{eq:bianchiID-1}
imposes the scaling relation $k^{2}=\xi\,R$, with $\xi=\frac{1}{4\lambda_{\ast}}$.
The RG running~\eqref{eq:scalingUV} thus yields an effective gravitational
action of the form~\cite{Platania:2019qvo}
\begin{equation}
S_{grav}^{\text{{eff}}}=\int d^{4}x\sqrt{-g}\;\left\{ \frac{R^{2}}{128\pi g_{\ast}\lambda_{\ast}}+f_{RG}(R)\right\} \,,\label{eq:actionJF}
\end{equation}
where $f_{RG}(R)$ is the part of the action generated as the RG trajectories
flow away from the NGFP
\begin{equation}
f_{RG}(R)=b_{1}R^{\frac{4-\theta_{1}-\theta_{2}}{2}}+b_{2}R^{\frac{4-\theta_{1}}{2}}+b_{3}R^{\frac{4-\theta_{2}}{2}}+b_{4}R^{2-\theta_{1}}+b_{5}R^{2-\theta_{2}}\,,\label{eq:eff2}
\end{equation}
with the coefficients $b_{i}$ being defined by
\begin{equation} 
\begin{array}{ll}
b_{1}=\frac{c_{1} c_{2}\left(e_{1}^{1} e_{2}^{2}+e_{1}^{2} e_{2}^{1}\right)\left(4 \lambda_{*} M_{P l}^{2}\right)^{\frac{\theta_{1}+\theta_{2}}{2}}}{128 \pi\left(g_{*} \lambda_{*}\right)^{2}}, 
& \quad b_{2}=\frac{c_{1}\left(e_{1}^{2} \lambda_{*}-e_{1}^{1} g_{*}-2 e_{1}^{2} \lambda_{*}\right)\left(4 \lambda_{*} M_{P l}^{2}\right)^{\frac{\theta_{1}}{2}}}{128 \pi\left(g_{*} \lambda_{*}\right)^{2}}, \\ 
b_{3}=\frac{c_{2}\left(e_{2}^{2} \lambda_{*}-e_{2}^{1} g_{*}-2 e_{2}^{2} \lambda_{*}\right)\left(4 \lambda_{*} M_{P l}^{2}\right)^{\frac{\theta_{2}}{2}}}{128 \pi\left(g_{*} \lambda_{*}\right)^{2}}, 
& \quad b_{4}=\frac{c_{1}^{2}\left(e_{1}^{1} e_{1}^{2}\right)\left(4 \lambda_{*} M_{P l}^{2}\right)^{\theta_{1}}}{128 \pi\left(g_{*} \lambda_{*}\right)^{2}}, \\ 
b_{5}=\frac{c_{2}^{2}\left(e_{2}^{1} e_{2}^{2}\right)\left(4 \lambda_{*} M_{P l}^{2}\right)^{\theta_{2}}}{128 \pi\left(g_{*} \lambda_{*}\right)^{2}}. \end{array}
\end{equation}
Here $M_{Pl}$ is the reduced Planck mass, while the integration constants  $c_i$, the critical exponents $\theta_i$ and the eigenvectors $\mathbf{e}_i$ are those introduced in Sect. \ref{sect2b}, cf. Eq.~\eqref{eq:scalingUV}.

The gravitational action
\begin{equation}
S_{grav}^{\ast}=\int d^{4}x\sqrt{-g}\;\frac{R^{2}}{128\pi g_{\ast}\lambda_{\ast}}\label{eq:FPaction}
\end{equation}
is the fixed-point action~\cite{Domazet:2012tw,Platania:2019qvo}.
This is compatible with the results in~\cite{Benedetti:2012dx,Dietz:2012ic,Demmel:2015oqa}
where, using the FRG equation~\eqref{eq:floweq} to study the RG flow
of $f_{k}(R)$-gravity, it has been shown that the fixed-point Lagrangian
is $\mathcal{L}_{\ast}=f_{\ast}(R)\propto R^{2}$. Additional $R^{n}$-operators
are generated along the RG flow. The effective action in~\eqref{eq:actionJF}
is thus expected to capture key features of the gravitational RG flow:
at the fixed point the action is $S_{grav}^{\ast}$. Lowering the
RG scale $k$ down towards the infrared, additional operators are
generated and $\mathcal{L}_{grav}=\mathcal{L}_{\ast}+\delta\mathcal{L}_{RG}$.
The set of operators appearing in $\delta\mathcal{L}_{RG}$ depends
on how the RG trajectories emerge from the NGFP. In the simplified
model~\eqref{eq:actionJF}, this information lies in the critical
exponents $\theta_{i}$. In what follows we will explore the consequences
of this fact in inflationary cosmology~\cite{Bonanno:2018gck,Platania:2019qvo}.

Provided that $f_{RG}^{(2)}(R)\neq-\frac{1}{64\pi g_{\ast}\lambda_{\ast}}$,
the gravitational action~\eqref{eq:actionJF} is conformally equivalent
to Einstein gravity, minimally coupled with a scalar field $\phi$
\begin{equation}
S_{grav}=\int d^{4}x\sqrt{-g_{E}}\;\left(\frac{R_{E}}{16\pi G_{0}}+\frac{1}{2}g_{E}^{\mu\nu}\partial_{\mu}\phi\partial_{\nu}\phi-V(\phi)\right)\,,\label{eq:Eaction}
\end{equation}
where the subscript ``E'' indicates that these quantities are computed
using the metric $g_{\mu\nu}^{E}=e^{\sqrt{2/3}\phi/M_{Pl}}\,g_{\mu\nu}$
in the Einstein frame, and $V(\phi)=U(\varphi(\phi))\,\varphi(\phi)^{-2}$,
with
\begin{equation}
\varphi=16\pi G_{0}\left(\frac{R}{64\pi g_{\ast}\lambda_{\ast}}+f'_{RG}(R)\right)=e^{\sqrt{2/3}\phi/M_{Pl}}\,
\end{equation}
and
\begin{equation}
U(\varphi)=\frac{R[\varphi]^{2}}{128\pi g_{\ast}\lambda_{\ast}}-f_{RG}(R[\varphi])+R[\varphi]\;f'_{RG}(R[\varphi])\,.
\end{equation}
A period of exponential grow of the scale factor occurs if the dynamics
of the scalar field $\phi$ is dominated by its potential energy $V(\phi)$.
This happens under the slow-roll conditions
\begin{equation} \epsilon(\phi) \equiv \frac{1}{2 \kappa}\left(\frac{V^{\prime}(\phi)}{V(\phi)}\right)^{2}\ll 1\,, \quad \eta(\phi) \equiv \frac{1}{\kappa}\left(\frac{V^{\prime \prime}(\phi)}{V(\phi)}\right)\ll 1 \, .
\end{equation}
The violation of the slow-roll conditions, encoded in the equation
$\epsilon(\phi_{f})=1$, defines the value of the field at the end
of inflation, $\phi_{f}\equiv\phi(t_{f})$. The initial condition
$\phi_{i}\equiv\phi(t_{i})$ is then obtained by fixing the number
of e-folds
\begin{equation}
N(\phi_{i})=\int_{\phi_{f}}^{\phi_{i}}\frac{V(\phi)}{V'(\phi)}d\phi
\end{equation}
before the end of inflation. In the slow-roll approximation, the spectral
index and tensor-to-scalar ratio characterizing the scalar power spectrum
$\mathcal{P}_{s}(k)$ in Eq.~\eqref{eq:Sps} can be easily computed
by means of the following relations~\cite{Baumann:2009ds}
\begin{equation}
n_{s}=1-6\,\epsilon(\phi_{i})+2\,\eta(\phi_{i})\,,\qquad r=16\,\epsilon(\phi_{i})\,.\label{eq:nsr}
\end{equation}
Moreover, every inflationary model has to be ``normalized''~\cite{inflationaris},
i.e., the inflaton mass has to be fixed by requiring that the amplitude
$A_{s}$ of the scalar power spectrum~\eqref{eq:Sps} is
\begin{equation}
A_{s}=\frac{V(\phi_{i})}{24\pi^{2}M_{Pl}^{4}\;\epsilon(\phi_{i})}\simeq2.2\cdot10^{-19}\;.
\end{equation}
At the NGFP ($k\to\infty$) $f_{RG}(R)=0$ and the effective action~\eqref{eq:actionJF} reduces to the fixed-point action~$S_{grav}^{\ast}$.
In the Einstein frame, the corresponding fixed-point scalar potential
is constant
\begin{equation}
V_{*}(\phi)=8\pi g_{\ast}\lambda_{\ast}M_{Pl}^{4}\,,\label{eq:flatpotential}
\end{equation}
and therefore it would generate an exactly scale-invariant power spectrum,
with $n_{s}=1$. This is compatible with the discussion made in Sect.~\ref{subsec:asinfl} and based on the scaling of the background graviton
propagator at the NGFP~\cite{Lauscher:2002sq,Bonanno:2002zb,Lauscher:2005qz,Bonanno:2007wg}.
The mass scale associated with the scalar degree of freedom $\phi$
can be read off from the the potential $V_{\ast}(\phi)$ and reads
\begin{equation}
m^{2}=8\pi\left(\frac{4}{3}\,\lambda_{\ast}g_{\ast}\right)M_{Pl}^{2}\,.\label{eq:infmass}
\end{equation}
This mass depends on fixed-point quantities only via the universal
product $(\lambda_{\ast}g_{\ast})$~\cite{Lauscher:2001ya}, as expected
from the universality properties of the theory at the NGFP.

In the model~\eqref{eq:actionJF} the departure from the exact scale
invariance is due to the departure of the RG flow from the NGFP. Lowering
the RG scale down towards the infrared, the gravitational Lagrangian
is modified by the operators in $\delta\mathcal{L}=f_{RG}(R)$ and,
in the Einstein frame, this corresponds to a variation of the scalar
potential $V(\phi)$,
\begin{equation}
V_{\ast}\to V(\phi)=V_{\ast}+\delta V(\phi)\,.
\end{equation}
Its form is determined by the critical exponents $\theta_{i}$, which
are real numbers in the case of the most commonly studied gravity-matter
systems~\cite{Dona:2013qba,Biemans:2017zca,Alkofer:2018fxj}. The
asymptotic-safety condition requires the real part of the critical
exponents to be positive, $\mathrm{Re}(\theta_{i})>0$. As we are
interested in the case of gravity-matter systems, we will only focus
on the case where the critical exponents are real. It is assumed,
in a first approximation, that the energy-density of the inflaton
field $\phi$ dominates. Under this assumption, the other matter fields
do not contribute to the inflationary dynamics~\cite{Bonanno:2018gck}.

\begin{figure}[t!]
\begin{centering}
\includegraphics[scale=0.33]{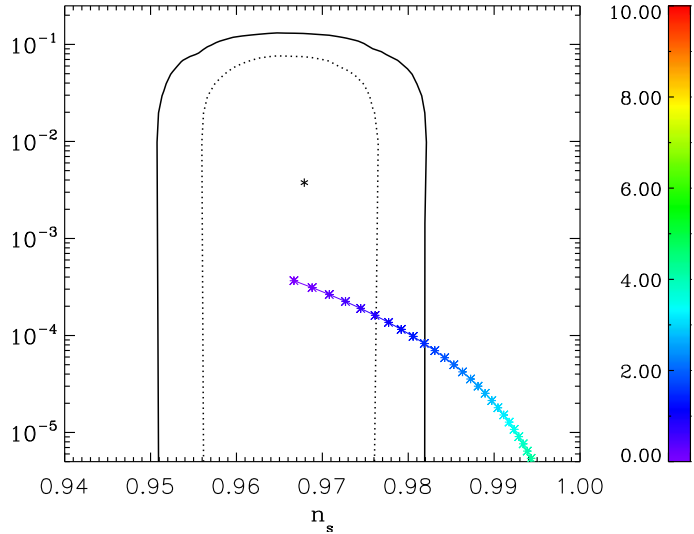}
\par\end{centering}
\caption{Spectral index and tensor-to-scalar ratio induced by the family of
theories in Eq.~\eqref{eq:badtheories} as a function of the power
index $p$ for $N=60$ e-folds and $b=1$. Dashed and solid lines
correspond to the $1\sigma$ and $2\sigma$ confidence levels on the
values of $(n_{s},r)$ extracted from the Planck data~\cite{Akrami:2018odb}.
The Starobinsky model, denoted with a star symbol, is also shown for
comparison. Only theories with $p\protect\leq1$ are reasonably within
the $2\sigma$ confidence-level line.}
\label{fig:minusplag}
\end{figure}

It is crucial to note that if all critical exponents are $\theta_{i}>4$,
the effective Lagrangian $\mathcal{L}_{\mathrm{eff}}$ reads
\begin{equation}
\mathcal{L}_{\mathrm{eff}}=\mathcal{L}_{*}+b\,R^{-p},\quad p>0\,,\label{eq:badtheories}
\end{equation}
where $R^{-p}$ is the dominant correction in $f_{RG}(R)$ and $b$
is the corresponding coefficient. Contributions of the form $R^{-p}$
are suppressed when $R$ is large, so that the deviation from the
exact scale invariance would be negligible (see Fig.~\ref{fig:minusplag}).
Moreover in this case the the $R$-operator is not generated by the
flow. In the case $\theta_{i}=4$, the model~\eqref{eq:actionJF}
gives rise to an inflationary scenario compatible with the Planck
data only under specific conditions~\cite{Platania:2019qvo}.
The agreement with the Planck data thus requires that at least one
of the critical exponents is $\theta_{i}<4$. This condition is realized,
e.g., when gravity is minimally-coupled to the fields of the Standard
Model, at least in the approximation where these fields are free~\cite{Dona:2013qba,Biemans:2017zca,Alkofer:2018fxj}.
Within the simple model reviewed here, matter models making all gravitational
critical exponents $\theta_{i}>4$ would not be compatible with observational
data. In this sense, the Planck data on the CMB anisotropies could
be used to constrain the primordial matter content of the universe
\cite{Bonanno:2018gck}.

\begin{figure}[t!]
\centering{}\includegraphics[width=0.75\textwidth]{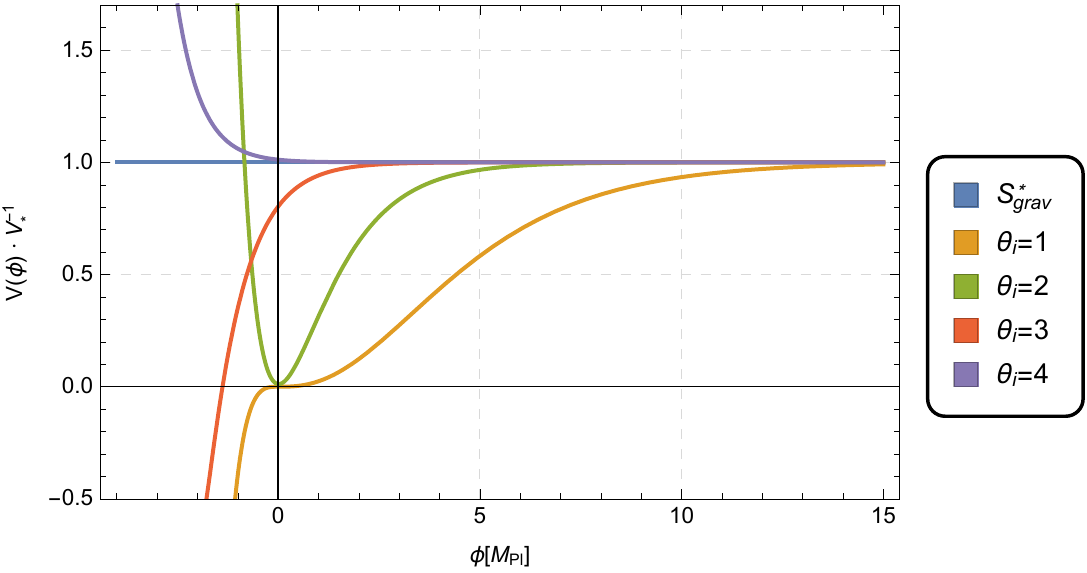}\caption{Inflationary potentials $V(\phi)$ produced by the conformal transformation
of the action~\eqref{eq:actionJF} for various values of the critical
exponents $\theta_{1}=\theta_{2}$~\cite{Platania:2019qvo}. 
The fixed-point
potential $V_{\ast}=8\pi g_{\ast}\lambda_{\ast}M_{Pl}^{4}$ associated
with the fixed-point action $S_{grav}^{\ast}$, Eq.~\eqref{eq:FPaction},
is also shown for comparison. It corresponds to an exactly scale invariant
scalar power spectrum, $n_{s}=1$. When the RG flow departs from the
NGFP, additional operators are generated by the flow. These operators
break the perfect scale invariance realized at the NGFP and destabilize
the fixed-point potential, $V_{\ast}\to V(\phi)=V_{\ast}+\delta V(\phi)$,
such that $V'(\phi)\protect\neq0$ at $\phi\sim M_{Pl}$. The scalar
field $\phi$ thus acquires a RG-induced kinetic energy, $\dot{\phi}_{i}\sim-V'(\phi_{i})/3H(t_{i})$.
The subsequent dynamics depends crucially on the critical exponents
$\theta_{i}$. In particular, the case $\theta_{1}=\theta_{2}=2$
reproduces the well-known Starobinsky model.\label{Fig1}}
\end{figure}

The scalar potential $V(\phi)$ is shown in Fig.~\ref{Fig1}
for various values of $\theta_{1}=\theta_{2}$. All functions $V(\phi)$
approach the same constant value $V_{\ast}$ for $\phi\gg M_{Pl}$.
In fact, as soon as $\theta_{i}\neq0$, the coupling of the $R^{2}$-term
in Eq.~\eqref{eq:actionJF} is not modified by the presence of the
additional operators in $\delta\mathcal{L}=f_{RG}(R)$. Therefore,
the height of the plateau and the inflaton mass are those in Eq.~\eqref{eq:flatpotential} and Eq.~\eqref{eq:infmass}. This is an artifact of the simplified
model in~\cite{Bonanno:2018gck}: the coupling to the $R^{2}$
operator is a running quantity and therefore also the value of the
scalar potential at $\phi\gg M_{Pl}$ should vary along the flow.
In other words, in a more elaborate model accounting for the running
of the coupling $B$ in Eq.~\eqref{eq:quadgrav}, the family of effective
potentials $V(\phi)$ should be characterized by a plateau with~$V_{plateau}\neq V_{\ast}$:
this decoupling would allow to set the initial conditions for inflation
at Planckian scales and, at the same time, to reproduce the correct
amplitude of scalar perturbations at the horizon exit~\cite{Platania:2019qvo}.
As the RG flow moves way from the NGFP, the scalar potential $V(\phi)$
is dynamically modified such that $V'(\phi)$ is generally non-zero.
The scalar field $\phi$ thus acquires a RG-running-induced kinetic
energy $\dot{\phi}_{i}\sim-V'(\phi_{i})/3H(t_{i})$. This provides
the initial conditions for the subsequent evolution of the scale factor
$a(t)$, according to the modified Friedmann equations. Depending
on the RG-induced variation of the scalar potential $\delta V(\phi)$,
the dynamics of the scalar field $\phi$ can trigger a period of slow-roll
inflation. For instance, the case $\theta_{1}=\theta_{2}=2$ gives
the scalar potential
\begin{equation}
V(\phi)=e^{-2\sqrt{\frac{2}{3}}\frac{\phi}{M_{Pl}}}\left\{ \frac{3}{4}\left(1-e^{\sqrt{\frac{2}{3}}\frac{\phi}{M_{Pl}}}\right)^{2}m^{2}+\Lambda_{\mathrm{{eff}}}\right\} M_{Pl}^{2}\,,\label{eq:theta2}
\end{equation}
i.e., a Starobinsky-like potential in the presence of an effective
cosmological constant $\Lambda_{\mathrm{{eff}}}=-(b_{1}+b_{4}+b_{5})M_{Pl}^{2}$
(see Fig.~\ref{Fig1}). As it is well known, this model leads
to cosmic parameters
\begin{equation}
\begin{split}
&n_{s}\simeq1-\frac{2}{N}+\mathcal{O}(N^{-3})\,,\\[0.1cm]
&r\simeq\frac{12}{N^{2}}+\mathcal{O}(N^{-3})\,,\\
\end{split}
\end{equation}
in good agreement with the current observational data. In the case
$\theta_{1}\approx2$ and $\theta_{2}\approx4$, realized when gravity
is minimally coupled with the (free) matter fields of the Standard
Model~\cite{Biemans:2017zca,Alkofer:2018fxj}, the action~\eqref{eq:actionJF}
differs from the one of the Starobinsky model by subleading terms
of the form $R^{-p}$, with $p>0$, which are suppressed for large
$R$. The inflationary dynamics, as shown in~\cite{Bonanno:2018gck},
is thus very similar to that of the Starobinsky model (case $\theta_{1}=\theta_{2}=2$).
In the next subsection we will see how this scenario is modified when
the RG-improved effective action is obtained by starting from the
quadratic gravity action~\eqref{eq:quadgrav}.

\subsection{Comparison with the Planck data in RG-improved quadratic gravity}

In~\cite{Bonanno:2015fga,Bonanno:2016rpx} a class of inflationary
models arising from the RG improvement of quadratic gravity (without
matter) has been investigated (see also~\cite{Bonanno:2012jy,Hindmarsh:2012rc}).
According to the studies of the non-perturbative RG flow of truncated
$f(R)$-theories without matter~\cite{Lauscher:2002sq,Rechenberger:2012pm},
the NGFP is attractive with respect to three relevant directions,
those associated with the dimensionless couplings $(g_{k},\lambda_{k},\beta_{k})$,
with $\beta_{k}=B(k)/G(k)$~\cite{Falls:2014tra,Falls:2017lst,Falls:2018ylp,Alkofer:2018fxj}.
The question motivating the studies in~\cite{Bonanno:2015fga,Bonanno:2016rpx}
is whether the scale dependence of all relevant gravitational couplings
can modify the classical Starobinsky model and if the RG-improved
model is compatible with the Planck data. The ansatz for the gravitational
action is
\begin{equation}
\mathcal{L}_{k}=\frac{1}{16\pi g_{k}}\left(R-2\lambda_{k}k^{2}\right)-\beta_{k}R^{2}\,.
\end{equation}
In order to derive an analytical form for the inflationary potential
in the Einstein frame, the running of the gravitational couplings
is approximated by~\cite{Bonanno:2015fga,Bonanno:2016rpx}
\begin{equation}
g_{k}=\frac{(c_{1}\mu^{-2})k^{2}}{1+g_{\ast}^{-1}(k^{2}-\mu^{2})(c_{1}\mu^{-2})}\,,\qquad\beta_{k}=\beta_{\ast}+b_{0}\left(\frac{k^{2}}{\mu^{2}}\right)^{-1/2}\,,\qquad\lambda_{k}=c_{2}k^{-2}\,,
\end{equation}
where $\mu$ is a reference scale and the three parameters $(b_{0},c_{1},c_{2})$,
corresponding to the three relevant directions of the theory, identify
the RG trajectories terminating at the NGFP in the ultraviolet limit.
These are free parameters of the theory and must be fixed by comparing
the results with observations. Using the cutoff $k^{2}=\xi R$~\cite{Bonanno:2015fga,Bonanno:2016rpx},
the RG-improved effective action reads
\begin{equation}
S_{grav}^{\text{{eff}}}=\frac{1}{2\kappa}\int d^{4}x\sqrt{-g}\left\{ R-2\tilde{\Lambda}+\frac{R^{2}}{6m^{2}}+\tilde{\alpha}R^{3/2}\right\} \,,\label{eq:actstarimp}
\end{equation}
with
\begin{equation} \begin{array}{ll}
{\kappa=\frac{48 \pi^{2} c_{1}}{6 \pi \mu^{2}-23\left(\mu^{2}+2 \xi c_{2}\right) c_{1}},} 
& \quad {m^{2}=\frac{8 \pi^{2}}{\kappa\left(23 \xi-96 \pi^{2} \beta_{*}\right)}} \,, \\[0.3cm] {\tilde{\Lambda}=\frac{\mu^{2}\left(6 \pi-23 c_{1}\right) c_{2}}{6 \pi \mu^{2}-23\left(\mu^{2}+2 \xi c_{2}\right) c_{1}},} 
& \quad {\tilde{\alpha}=-2 b_{0} \kappa\left(\xi \mu^{-2}\right)^{-\frac{1}{2}}} \,.\end{array} \end{equation}
The inflationary scenario generated in this model can be studied in
the Einstein frame, where the $f(R)$ action~\eqref{eq:actstarimp}
can be written as in Eq.~\eqref{eq:Eaction}. In this case the scalar
potential $V(\phi)$ reads~\cite{Bonanno:2015fga,Bonanno:2016rpx}
\begin{equation} \begin{array}{l} {V_{\pm}(\phi)=\frac{m^{2} e^{-2 \sqrt{\frac{2 \kappa}{3} \phi}}}{256 \kappa}\left\{192\left(e^{\sqrt{\frac{2 \kappa}{3}} \phi}-1\right)^{2}-3 \alpha^{4}+128 \Lambda\right.} \\ {-\sqrt{32} \alpha\left[\left(\alpha^{2}+8 e^{\sqrt{\frac{2 \kappa}{3}} \phi}-8\right) \pm \alpha \sqrt{\alpha^{2}+16 e^{\sqrt{\frac{2 \kappa}{3} \phi}}-16}\right]^{\frac{3}{2}}} \\ {\left.-3 \alpha^{2}\left(\alpha^{2}+16 e^{\sqrt{\frac{2 \kappa}{3}} \phi}-16\right) \mp 6 \alpha^{3} \sqrt{\alpha^{2}+16 e^{\sqrt{\frac{2 \kappa}{3}} \phi}-16}\right\}} \,, \end{array} \end{equation}
with the dimensionless couplings $\Lambda$ and $\alpha$ given by
$\Lambda=m^{-2}\tilde{\Lambda}$ and $\alpha=3\sqrt{3}m\tilde{\alpha}$.
The existence of two solutions is due to the presence of the additional
$R^{3/2}$-term in the effective action, and the standard Starobinsky
model is recovered by setting $\alpha=\Lambda=0$. Both functions
$V_{\pm}(\phi)$ define a two-parameters family of potentials, parametrized
by the couple $(\alpha,\Lambda)$. The common feature of these potentials
is the existence of a plateau for large positive values of the field
$\phi$, with $V_{plateau}=\frac{3m^{2}}{4\kappa}$. Note that the
inclusion of the running of the coupling $\beta_{k}$ now allows for
an effective potential with $V_{plateau}\neq V_{\ast}$, as mentioned
in the previous section.\\

\begin{figure}[t!]
\begin{centering}
\includegraphics[scale=0.75]{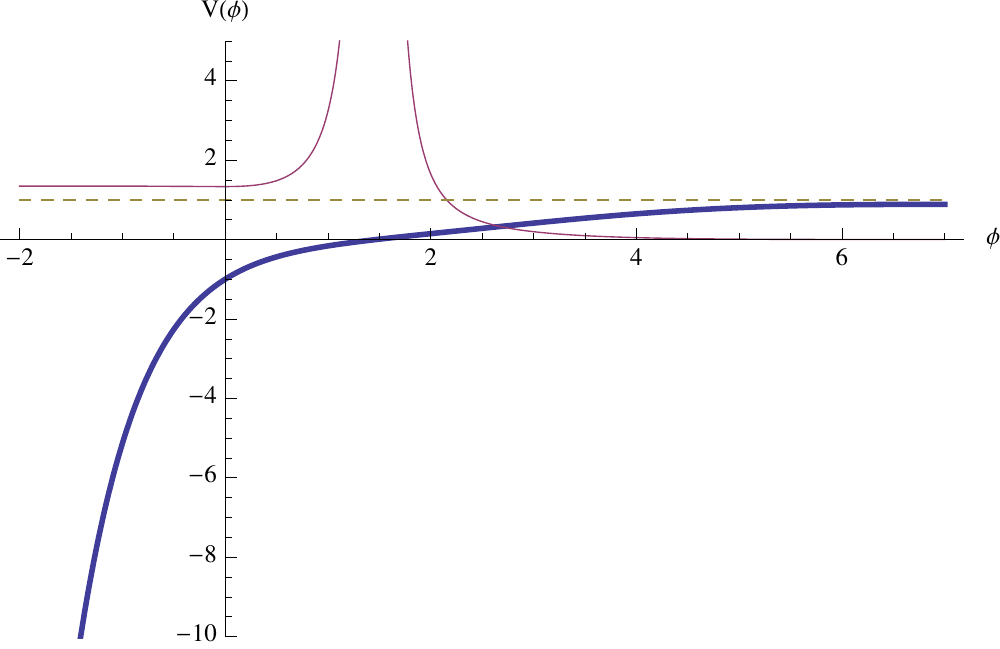}$\quad$\includegraphics[scale=0.75]{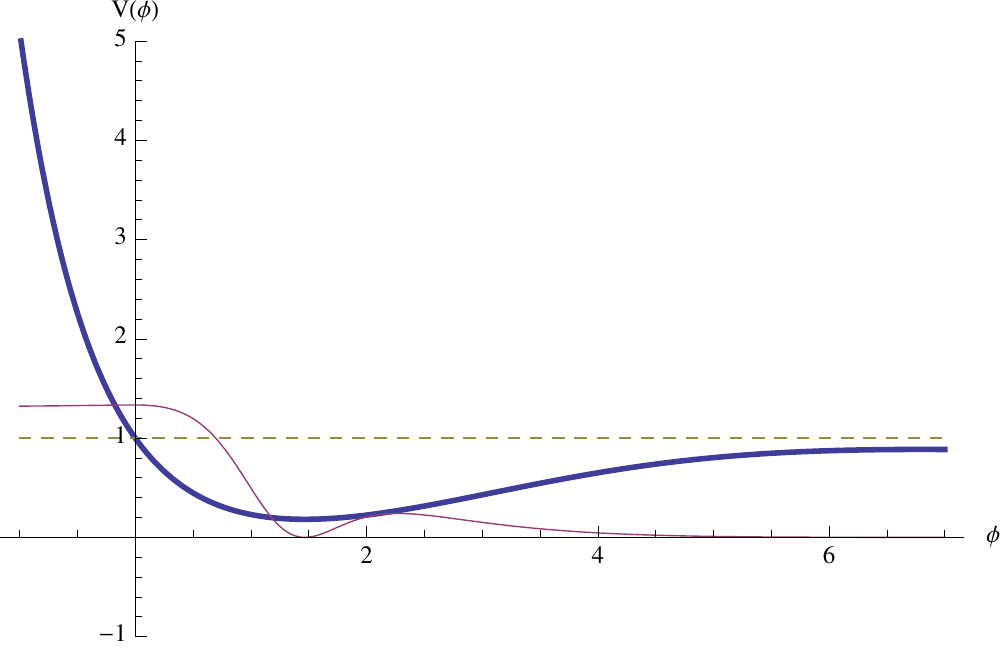}
\par\end{centering}
\caption{Scalar potential $V_{-}(\phi)$ (blue thick line) and slow-roll function
$\epsilon(\phi)$ (red thin line) for $\alpha=-10$ and $\Lambda=-2$
(potential unbounded from below, left panel) and $\Lambda=2$ (potential
with a minimum and $V_{\mathrm{min}}>0$, right panel)~\cite{Bonanno:2015fga,Bonanno:2016rpx}.
The slow-roll conditions are violated and inflation ends if $\exists\,t_{f}$
such that $\epsilon(\phi(t_{f}))=1$ (thin dashed line) and $\epsilon(\phi)>1$
for any $t>t_{f}$. The dynamics induced by the potential $V_{-}(\phi)$
in the right panel keeps the dynamical field $\phi(t)$ in the region
where $\epsilon(\phi)<1$~\cite{Bonanno:2015fga,Bonanno:2016rpx},
thus making it impossible to exit inflation by violation of the slow-roll
conditions. The potential plotted on the left panel allows instead
for a finite period of slow-roll inflation, but in this case the reheating
of the universe after inflation cannot be described via the standard
parametric oscillations of the inflaton field.}
\label{fig:potVminus}
\end{figure}

In order to fulfill the slow-roll conditions, the dynamical evolution
of the inflaton field must start from a quasi-deSitter state at $V(\phi_{i})\sim V_{\mathrm{plateau}}$,
and then proceed towards $\phi\ll M_{Pl}$. The inflationary dynamics
depends on the values of $(\alpha,\Lambda)$. For any $(\alpha,\Lambda)$,
the potential $V_{\pm}(\phi)$ can either develop a minimum (Fig.~\ref{fig:potVminus}, right panel) or be unbounded from below
(Fig.~\ref{fig:potVminus}, left panel). A standard reheating
phase is only possible in the first case. In addition, in these models
a ``graceful exist'' from inflation by violation of the slow-roll
conditions is only possible when $V_{\mathrm{min}}\leq0$. The case
$V_{\mathrm{min}}>0$ leads instead to eternal inflation, as shown
in the right panel of Fig.~\ref{fig:potVminus}.\\

\begin{figure}[t!]
\centering{}\includegraphics[scale=0.85]{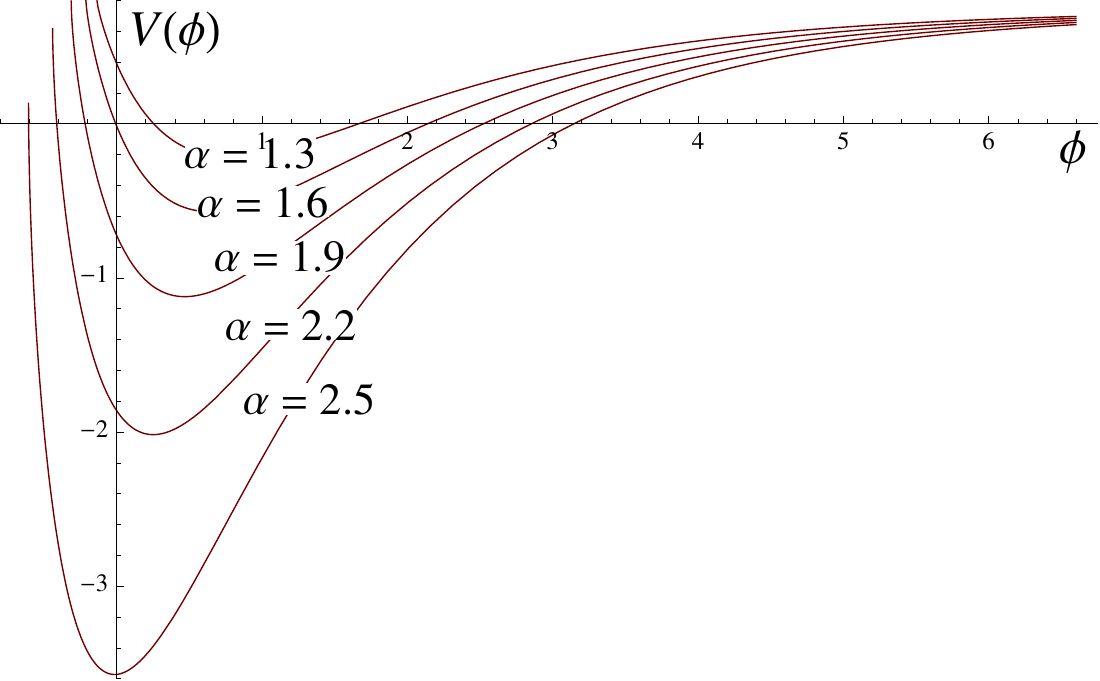}\caption{This figure depicts the potential $V_{+}(\phi)$ for $\Lambda=1.4$
and various values of $\alpha$ in the physically-interesting range,
$\alpha\in[1,3]$~\cite{Bonanno:2015fga,Bonanno:2016rpx}. This is
the class of potentials allowing for a (finite) period of slow-roll
inflation, followed by a standard reheating phase.}\vspace{-1cm} 
\label{fig:potentialplus}
\end{figure}

We now focus on the class of inflationary potentials providing a well
defined exit from inflation by violation of the slow-roll conditions,
followed by a phase of parametric oscillations of the inflaton field
\cite{Bonanno:2015fga,Bonanno:2016rpx}. These conditions are realized
by the class of potentials $V_{+}(\phi)$, with $\alpha\in[1,3]$
and $\Lambda\in[0,1.5]$~\cite{Bonanno:2015fga,Bonanno:2016rpx}.
The corresponding potential is shown in Fig.~\ref{fig:potentialplus}
for $\Lambda=1.4$ and various values of the effective coupling $\alpha$.
As already mentioned, the constants $(\alpha,\Lambda)$ parametrize
the deviations from the Starobinsky model due to the RG running of
the gravitational couplings. It is thereby interesting to understand
whether these modifications can affect the form of the power spectrum
of temperature fluctuations in the CMB, and if the values of the spectral
index $n_{s}$ and tensor-to-scalar ratio $r$ are modified by this
running. 

\begin{table}[t] \vspace{1.5cm} 	\renewcommand{\arraystretch}{1.4} 	
\begin{center} \begin{tabular}{|c|c||c|c||c|c||c|c|} \hline\hline \multicolumn{2}{|c||}{ \text{Cases}}  & \multicolumn{2}{|c||}{$N=50$} & \multicolumn{2}{|c||}{$N=55$}   & \multicolumn{2}{|c|}{$N=60$} \\ \hline\hline $\,\,\Lambda\,\,$ & $\,\,\,\alpha\,\,\,$ & $\quad n_s \quad$ & $ \quad \;\; r \;\;\quad $ & $\quad n_s \quad$ & $ \quad \;\;r \;\; \quad$ & $\quad n_s \quad$ & $\quad \;\; r \;\;\quad$ \\  \hline\hline  & 1.0 & 0.965 & 0.0069 & 0.968 & 0.0058 & 0.971 & 0.0050 \\  \cline{2-8} 0  & 1.8 & 0.966 & 0.0074 & 0.969 & 0.0063 & 0.972 & 0.0055 \\ \cline{2-8}  & 2.6 & 0.967 & 0.0076 & 0.969 & 0.0065 & 0.972 & 0.0056 \\ \hline\hline   & 1.0 & 0.965 & 0.0070 & 0.968 & 0.0059 & 0.971 & 0.0051 \\ \cline{2-8} 1  & 1.8 & 0.966 & 0.0074 & 0.969 & 0.0063 & 0.972 & 0.0055 \\ \cline{2-8}  & 2.6 & 0.967 & 0.0076 & 0.969 & 0.0065 & 0.972 & 0.0056 \\ \hline\hline \end{tabular} 
\caption{Values of the spectral index $n_s$ and tensor-to-scalar ratio $r$ obtained from the RG-improved inflationary model~\eqref{eq:actstarimp} for different values of $(\alpha,\Lambda)$ and number of e-folds $N$ in the range $50-60$~\cite{Bonanno:2016rpx}. The range of values obtained for the spectral index, $n_s\in[0.965,0.972]$, is in agreement with the one obtained by the Planck Collaboration, $n_s=0.968\pm0.006$, and the tensor-to-scalar ratio is always compatible with their upper limit, $r<0.11$. \label{TabPla1}} \end{center}\end{table}
The results are summarized in Tab.~\ref{TabPla1}
\cite{Bonanno:2016rpx}, where the values of $n_s$ and $r$ are shown for different values of the effective couplings of $(\alpha,\Lambda)$ and number of e-folds $N$. The range of values for the
spectral index is~$n_{s}\in[0.965,0.972]$~\cite{Bonanno:2016rpx}, in agreement with the
value extracted from the Planck data, $n_{s}=0.968\pm0.006$ ~\cite{Akrami:2018odb}. Moreover, the
tensor-to-scalar ratio is always compatible with their upper limit, $r<0.11$,
but it is slightly higher than the one predicted within the Starobinsky
model.

\section{Discussion\label{sec:7}}

The phenomenological consequences of Asymptotically Safe Gravity (ASG) are typically investigated within models that take the running
of the gravitational couplings into account. Based on the decoupling
mechanism~\cite{Reuter:2003ca}, it is expected that these models
can provide a qualitative, yet simple and intuitive, understanding
of the effective modifications of General Relativity induced by quantum
gravity in the asymptotic-safety approach. Nonethless, the derivation of these ``renormalization group (RG) improved'' models is not free of ambiguities. Bearing in mind strenghts and limitations of this approach, the scope of this review was to provide an overview of the main cosmological implications of ASG derived from models of RG-improved cosmology.

ASG is based on the existence of an interacting fixed point which
is attained by the gravitational RG flow in
the ultraviolet limit. The scale invariance of gravity at high energies
and the consequent gravitational antiscreening can be regarded as
the hallmarks of ASG. In particular the antiscreening character of
gravity, rendering the gravitational interaction weaker at high energies,
could lead to non-singular cosmological solutions: the classical singularity
could be replaced by a bounce or an emergent universe~\cite{Bonanno:2017gji}.
The entropy production during inflation might be attributed to an
energy flow from the gravitational to the matter degrees of freedom,
which causes the primordial evolution of the universe to be approximately
(but not exactly) adiabatic~\cite{Bonanno:2007wg}. Moreover, the
existence of a regime where gravity is approximately scale invariant
(fixed-point regime and departure of the RG flow from it) provides
a simple and natural interpretation for the nearly-scale-invariance
of the power spectrum of temperature fluctuations in the Cosmic Microwave
Background (CMB) radiation~\cite{Bonanno:2007wg}. In the case of
pure gravity, the spectral index and tensor-to-scalar ratio evaluated
from models of ``RG-improved'' inflation are in agreement with the
Planck data on CMB anisotropies~\cite{Bonanno:2015fga}. When matter
is minimally coupled to gravity, the universality properties of the
gravitational RG flow are modified and this modification depends on
the number and type (scalar, Dirac, vector, etc.) of matter fields:
the observational data could thus be used to constrain the matter
content of the theory in the early universe~\cite{Bonanno:2018gck}.
Finally, the running of the $R^{2}$-coupling could also provide a
mechanism to set the initial condition for inflation at trans-planckian
scales, while being able to reproduce the amplitude of the scalar
power spectrum at the horizon exit~\cite{Platania:2019qvo}. If this
mechanism is realized, it could provide a solution to the ``unlikeness
problem''~\cite{Ijjas:2013vea} of inflationary cosmology. A definite answer requires however more elaborate and extended studies, going beyond the simple models reviewed here.

Most of the results listed above have been obtained by including
the running of the gravitational couplings within the Einstein-Hilbert
truncation. As argued in~\cite{Lehners:2019ibe}, fourth-derivatives
operators are crucial for the understanding of the early-universe
evolution. The inclusion of these operators in the models of RG-improved
cosmologies could then be important to determine the phenomenological
implications of the gravitational antiscreening.

One of the main problems of models of ``RG-improved cosmologies''
is the identification of the physical cutoff acting as a decoupling
scale~\cite{Reuter:2003ca} for the RG flow of the effective average
action~\cite{Wetterich:1992yh}. The symmetries of
theory play an important role, as they could provide a guideline for
this scale-setting~\cite{Babic:2004ev,Domazet:2010bk,Koch:2010nn,Domazet:2012tw,Contreras:2016mdt}.
As shown in~\cite{Reuter:2004nv,Reuter:2004nx,Babic:2004ev,Domazet:2012tw,Koch:2014joa},
the contracted Bianchi identities typically lead to a ``consistency condition''
which can be used to determine the form of the
cutoff scale. Close to a fixed point, the RG flow should be universal
(no scheme or regulator dependence, at least for the flow of the full
-- not truncated -- effective action) and all physical scales collapse
into one: in this case the scale-setting is essentially unique. Away
from the fixed point, the physical cutoff is dictated by the consistency
condition. However, this condition requires the running of the gravitational
couplings as an external input: since the RG flow obtained from the
Wetterich equation~\cite{Wetterich:1992yh} depends explicitly on
the choice of the regulator, this dependence is inherited by the physical cutoff scale and it disappears only in the proximity of the fixed point (provided that no truncation of the effective action is employed).

While it is expected that the RG-improved models capture the qualitative
features of the quantum modifications of General Relativity according
to ASG, quantitative results require the knowledge of the fully-quantum
gravitational effective action. This is expected to be non-local,
due to the resummation of quantum fluctuations on all scales. 
Progress in this direction has been made in \cite{Codello:2015mba}, where the leading-order, quadratic part of the effective action has been derived within the framework of effective field theory, in~\cite{Codello:2015oqa}, by studying the flow of the non-local part of the one-loop effective action, and, more recently, in~\cite{Knorr:2019atm}, where the beta functions for a specific non-local action have been computed using the functional renormalization group. Future developments of these programs could provide indications
on the form of the gravitational effective action. On the phenomenological
side, this could allow to derive more quantitative results on the
implications of ASG in astrophysics and cosmology.

\section*{Conflict of Interest Statement}
The authors declare that the research was conducted in the absence of any commercial or financial relationships that could be construed as a potential conflict of interest.

\section*{Acknowledgments}
{We acknowledge financial support by the Baden-Württemberg Ministry of Science, Research and the Arts and by Ruprecht-Karls-Universität Heidelberg. The author also thanks A. Bonanno and A. Eichhorn for their comments on this manuscript. The research of AP is supported by the Alexander von Humboldt Foundation.}

\bibliographystyle{elsarticle-num}
\bibliography{AleBib}

\end{document}